\documentclass[journal, twocolumn]{IEEEtran}
\usepackage[T1]{fontenc}
\usepackage{cite}
\usepackage{amsmath,amssymb}
\usepackage[ruled,vlined]{algorithm2e}
\usepackage{algorithmic}
\usepackage{graphicx}
\usepackage{subfigure}
\usepackage{textcomp}
\usepackage{xcolor}
\usepackage{color}
\usepackage{url}
\usepackage{bm}
\usepackage{diagbox}
\usepackage{newtxmath}
\graphicspath{{./fig/}}
\newcommand{\qw}{{\bf w}}
\newcommand{\qa}{{\bf a}}

\newcommand{\qX}{{\bf X}}
\newcommand{\qB}{{\bf B}}
\newcommand{\qR}{{\bf R}}

\newcommand{\qQ}{{\bf Q}}

\newcommand{\qY}{{\bf Y}}
\newcommand{\qv}{{\bf v}}
\newcommand{\qN}{{\bf N}}
\newcommand{\qp}{{\bf p}}
\newcommand{\qq}{{\bf q}}
\newcommand{\qL}{{\bf L}}
\newcommand{\qM}{{\bf M}}
\newcommand{\qT}{{\bf T}}
\newcommand{\qh}{{\bf h}}
\newcommand{\qH}{{\bf H}}
\newcommand{\qI}{{\bf I}}
\newcommand{\qU}{{\bf U}}
\newcommand{\qg}{{\bf g}}
\newcommand{\qW}{{\bf W}}
\newcommand{\tr}{\mbox{trace}}
\newcommand{\be}{\begin{equation}} \newcommand{\ee}{\end{equation}}
\newcommand{\bea}{\begin{eqnarray}} \newcommand{\eea}{\end{eqnarray}}

\begin{document}
\title{Model-driven Learning for Generic MIMO Downlink Beamforming With Uplink Channel Information}
\author{Juping Zhang, Minglei You, Gan Zheng,~\IEEEmembership{Fellow,~IEEE,}
       Ioannis Krikidis,~\IEEEmembership{Fellow,~IEEE,}
       and Liqiang Zhao,~\IEEEmembership{Member, IEEE} 
\thanks{J. Zhang, M. You and G. Zheng are with the Wolfson School of Mechanical, Electrical and Manufacturing Engineering, Loughborough University, Loughborough, LE11 3TU, UK (E-mail: \{j.zhang3, m.you, g.zheng\}@lboro.ac.uk).}
  \thanks{I. Krikidis is with the    Department of Electrical and Computer Engineering,  University of Cyprus, 1678 Nicosia, Cyprus (E-mail: krikidis@ucy.ac.cy).}
\thanks{L. Zhao is with the State Key Laboratory of Integrated Service Networks at Xidian University, Xi'an 710071, China (E-mail: lqzhao@mail.xidian.edu.cn).}
}

\markboth{\it A Manuscript Accepted in IEEE Transactions on Wireless Communications}{}
\maketitle
\vspace{-8mm}
\begin{abstract}
Accurate downlink channel information is crucial to the beamforming design, but it is difficult to obtain in practice. This paper investigates a deep learning-based optimization approach of the downlink beamforming to maximize the system sum rate, when only the uplink channel information is available. Our main  contribution is to propose a model-driven learning technique that exploits the structure of the optimal downlink beamforming to design an effective hybrid learning strategy with the aim to maximize the sum rate performance. This is achieved by jointly considering the learning performance of  the downlink channel, the power and the sum rate in the training stage.  The proposed approach applies to generic cases in which the uplink channel information is available, but its relation to the downlink channel is unknown and does not require an explicit downlink channel estimation. We further extend the developed technique   to   massive multiple-input multiple-output scenarios and achieve a  distributed learning strategy for multicell   systems without an inter-cell signalling overhead.  Simulation results verify that our proposed method   provides the performance close to the state of the art numerical algorithms with perfect downlink channel information and significantly outperforms existing data-driven methods in terms of the sum rate.
\end{abstract}

\vspace{-2mm}

\section{Introduction}
 Beamforming is an important multi-antenna technique to deal with interference and improve the capacity of multiuser wireless communications systems. Most beamforming design problems are nonconvex, so early efforts to optimize beamforming mainly refer to numerical algorithms  \cite{bengtsson1999optimal}\cite{ shi2011iteratively}, which cause  high complexity for practical implementation. Recently deep learning has been recognized as a new ``learn to optimize'' approach to design beamforming \cite{sun2017learning,alkhateeb2018deep,Jin19,huang2019fast,xia2019deep}. The deep learning approach significantly reduces the optimization complexity and the resulting beamforming solution can be potentially implemented in real time.

  The optimization and performance of beamforming critically depend  on the availability of perfect channel state information (CSI) at the transmitter. Existing deep learning based beamforming solutions are mainly based on the availability of perfect CSI.  In the downlink,   perfect CSI is difficult to obtain at the base station (BS) for several reasons. In the time division duplex (TDD) systems, normally channel reciprocity is assumed, so the downlink CSI can be estimated from the pilots sent by the mobile users in the uplink. However, in practice, the  channel reciprocity does not hold because the analog radio front-ends at the BS and the mobile users  exhibit non-reciprocity due to non-identical behavior of the individual transmit and receive chains \cite{reciprocity-tsp}. In the  frequency division duplex (FDD) systems, for the BS to obtain the CSI, the BS needs to  first transmit pilots  to the users for the downlink channel estimation, and then users feed back the estimated downlink channel to the BS. As a result, this channel acquisition process incurs a large overhead and reduces the effective system spectral efficiency.

  There has been significant efforts in obtaining  the CSI using deep learning techniques.  CsiNet was develop in \cite{CKWen2018v1} to  learn CSI sensing and recovery in FDD-based massive multiple-input and multiple-output (MIMO)  systems using the channel structure. A learned denoising-based approximate message passing   neural network was proposed in \cite{CKWen2018v2} for beamspace channel estimation in Millimeter-wave (mmWave) massive  MIMO systems.  
  Based on the fact that the uplink and downlink channels share the same propagation environment, deep neural network for channel calibration between the two directions was designed in \cite{Huang2019ULDL} for a generic massive MIMO system. A sparse complex-valued neural network was introduced in \cite{yang2019deep}  to approximate the uplink-to-downlink mapping function in FDD massive MIMO systems. Convolutional neural networks and generative adversarial networks were used in  \cite{UL2DL} to infer the downlink CSI by observing the uplink CSI.  The feasibility of channel mapping in space and frequency was demonstrated  in \cite{Alkhateeb2019-v2}, where the channels at one set of antennas using one frequency band can be learned from the channels at another set of antennas that use a different frequency band.   A comprehensive joint channel estimation and feedback framework based on deep learning  was proposed in \cite{joint_CE_Feedback},  which realizes the estimation, compression, and reconstruction of downlink channels in FDD massive MIMO systems.   There is an emerging direction of studies that aims to learn the beamforming solutions rather than improve the CSI accuracy, which is close to the main idea of this work. For instance,  the work in \cite{Feedback_massive} proposed a deep learning based CSI feedback framework for beamforming design in FDD massive MIMO systems to maximize the beamforming performance gain. A deep neural network using unsupervised training was proposed in \cite{Jiang2020IRS} to map the received uplink pilots to the beamforming matrix at the BS for the intelligent reflecting surface assuming uplink-downlink channel reciprocity. A channel sensing and hybrid precoding design was proposed in \cite{sensing_precoding}, by using the received pilots without the intermediate channel estimation step for TDD massive MIMO systems.

 While existing  deep learning approaches have achieved success in  individual tasks of channel estimation and beamforming design, they normally require massive amounts of data and computational resources, and their simple combination  will not guarantee satisfactory end results. This is  due to their inherent limitation of being data-driven and model-agnostic. The optimization of each separate task only focuses on its own objective assuming the other is ideal. In reality, the optimization in each task will introduce some error or imperfection, so the overall performance could deteriorate when they are simply putting together.

  A promising direction to remedy this problem is model-driven deep learning that combines the data-drive approach with the underlying domain knowledge, mathematical models  and problem structures, etc., to achieve a better inference  with less data. Recent advancements in model-driven deep learning approaches in physical layer communications were discussed in \cite{model_Li}. In our previous works \cite{xia2019deep}, \cite{Magazine_Zheng}, we have proposed the model-driven neural network design for beamforming optimization by exploiting the problem structure. Deep neural networks that adopt the algorithmic structure and constraints of adaptive signal processing techniques were proposed in \cite{adaptive_ultrasound} that can efficiently learn to perform fast high-quality ultrasound beamforming by using very few training data. Model-driving learning is a new concept that can be broadly applied in engineering design, and a comprehensive review of   leading approaches for combining model-based algorithms with deep learning can be found in \cite{model_learning}, with detailed signal processing and communications oriented examples.

 In this paper, we aim to design a model-drive deep learning approach to jointly tackle the challenge of channel estimation and optimize the downlink beamforming  to maximize the sum rate of generic multiuser MIMO systems. Different from the literature, we assume only information about the uplink channel is available, without explicit knowledge of the downlink CSI, and the relation between the downlink and uplink channels is unknown. The uplink channel information could take the form of either perfect CSI or received pilots. This method will alleviate the burden of channel estimation at the user side,  reduce the feedback overhead, and it is flexible enough to be used in both TDD and FDD systems, and in  massive MIMO and multicell settings. Especially for FDD systems, the proposed approach allows to learn the downlink beamforming directly without the need of sending downlink pilots, uplink feedback or explicit channel estimation.  The novelty of our work is two-fold: first,  we propose to optimize the beamforming in order to maximize the end performance   and therefore bypass the explicit intermediate channel estimation step; second, we introduce a model-driven deep learning-based approach. Comparing to existing data-driven  beamforming learning, our proposed approach specifies the most appropriate features to be learned with improved performance of inference and end performance.   Our main contributions are summarized as follows:
\begin{itemize}
\item  We exploit the algorithmic  structure of beamforming solutions as the useful model information, and  propose a hybrid  method for joint learning of the downlink channel and optimization of beamforming to guarantee the sum rate performance.  To be specific, we design a neural network consisting of two subnets for learning the downlink channel and the auxiliary power vector, respectively, from which the downlink beamforming solution can be constructed. The overall loss function is hybrid and chosen to be a weighted sum of the loss functions of channel and power learning using supervised training, and the sum rate using unsupervised training.
\item We investigate techniques to further reduce the  problem dimension and achieve near-optimal low-complexity learning, by  using the zero-forcing (ZF) beamforming in the loss function, which is specially appealing for massive MIMO systems.
\item We extend the proposed method to multicell massive MIMO systems in which a BS in each cell is able to learn the beamforming solution  in a distributed manner without signalling exchange with other cells.  To the best of our knowledge, this is the first distributed learning solution for the optimization of multicell beamforming.
\item  Extensive simulations are carried out to evaluate the performance the proposed algorithms, which show that the proposed algorithm can achieve a sum rate close to the weighted minimum mean squared error (WMMSE) algorithm \cite{shi2011iteratively}, and significantly outperforms existing learning methods.
\end{itemize}

The remainder of this paper is organized as follows. Section \ref{system_model} introduces the system model and the problem formulation. The uplink to downlink channel mapping is discussed in Section \ref{channel_mapping}. The model-driven hybrid learning approach for a general downlink  is proposed in section \ref{algorithm}. Section  \ref{massive}  presents techniques to further reduce the training complexity for single-cell massive MIMO systems. Section \ref{multicell} extends  the result to allow distributed learning of the beamforming solution in a multicell massive MIMO scenario. Simulation results and conclusions are given in Section \ref{simu} and Section \ref{conc}, respectively.

{\em Notions:} The boldface lower case letters and capital letters are used to represent column vectors and matrices, respectively. The notation $\mathbf{A}^H$ and $\|\mathbf{A}\|$ denote the transpose conjugate  and the  Frobenius-norm of a complex matrix $\mathbf{A}$, respectively. $\mathbb{C}$ denotes the complex field. The operator $\mathcal{CN}(\bf{m}, \mathbf{\Theta})$ represents a complex Gaussian vector with mean $\bf{m}$ and covariance matrix $\mathbf{\Theta}$. $\mathbf{I}_N$ denotes an $N\times N$ identity matrix. $E[\cdot]$ denotes the expectation of a random variable.

\section{System Model and Problem Formulation}\label{system_model}
 We start with a single-cell multi-input single-output (MISO) downlink system where a BS with $N_t$ antennas serves $K$ single-antenna users. The received signal at the user $k$ can be written as
\begin{align}\label{receivesing}
y_k=\mathbf{h}_{D,k}^H\mathbf{w}_ks_k+n_k,
\end{align}where $\mathbf{h}_{D,k}\in\mathbb{C}^{N_t\times 1}$ denotes the downlink channel vector from the BS to the user $k$,  $\mathbf{w}_k$ and $s_k$ denote the transmit beamforming vector and the information  signal for the user $k$ with normalized power, respectively. $n_k$ is the additive Gaussian white noise with zero mean and variance of $N_0$. The beamforming matrix is $\qW=[\qw_1, \cdots, \qw_K]$ and we collect the downlink CSI into $\qH_D=[\qh_{D,1}, \cdots, \qh_{D,K}]$. The received SINR  at user $k$ is expressed as
\begin{align}\label{sinr}
\gamma_k=\frac{|\mathbf{h}_{D,k}^H\mathbf{w}_k|^2}{\sum_{j=1, j\neq k}^K|\mathbf{h}_{D,k}^H\mathbf{w}_j|^2+N_0}.
\end{align}
The sum rate is then written as  $R^{\text{sum}}=\sum_{k=1}^K\log_2(1+\gamma_k)$.
Based on the above model, the sum rate maximization problem under the total transmit power constraint $P$ can be formulated as
\begin{align}\label{prob}
\max_{\qW} R^{\text{sum}},~~~~\mathrm{s.t.}~\sum_{k=1}^K\|\mathbf{w}_k\|^2\leq P.
\end{align}
The extended system models to   massive MIMO and multicell scenarios will be discussed in Sections \ref{massive} and \ref{multicell}, respectively, and specific techniques to reduce the training complexity and enable distributed learning will also be introduced.

 When the downlink CSI $\qH_D$ is available, there exist  numerical algorithms that can find the locally optimal beamforming solution of problem \eqref{prob} such as the WMMSE algorithm \cite{shi2011iteratively}. In our recent work \cite{xia2019deep}, we have proposed a deep learning method to solve this problem with perfect downlink CSI. However, without downlink CSI $\qH_D$, existing numerical or deep learning algorithms cannot be applied. Therefore, we focus on the design of downlink beamforming algorithms, when only the uplink channel information is available, either in the form of perfect CSI $\qH_U$  or the received pilot signal.

\section{Uplink to Downlink Channel Mapping}\label{channel_mapping}
In this paper, we rely on the uplink channel information to infer the downlink channel and optimize the downlink beamforming, so we assume that there exists a deterministic and  unique mapping from the uplink channel to the downlink channel but its explicit form is unknown and can be learned by a deep neural network. This assumption is based on the fact that a wireless channel between a transmitter-receiver pair is determined by the positions of the transmitter-receiver pair, antennas, carrier frequency and the environment in which the signals propagate  including the objectives and their materials and shapes within the environment. Because both uplink and downlink channels share the same   propagation environment, given the positions of the  transceiver pair, there is an intrinsic mapping between the uplink and the downlink channels. Below we will give details for the FDD and TDD cases, respectively.
\begin{itemize}
  \item In a FDD system, consider the single-antenna uplink channel $h_u$ and the downlink channel $h_d$ that operate at frequencies $f_u$ and $f_d$, respectively.  Assume that there are $N$ distinct propagation paths in the environment, the uplink channel can be written as
      \be
        h_u = \sum_{n=1}^N a_n e^{-j2\pi f_u \tau_n + j\phi_n},
      \ee
   where $a_n$ is the path attenuation, $\tau_n$ is the path delay  and $\phi_n$ is a frequency-independent phase shift that captures the reflection and attenuation effects of the signal along the path $n$.

   The path attenuation $a_n$ depends on the distance between the transceiver pair, their antenna gains, the carrier frequency and the environment,
   the phase shift $\phi_n$ depends on the scattering and the path delay $\tau_n$ depends on the propagation distance. Therefore when the environment and other factors are unchanged, there is a deterministic mapping from the positions to the channel \cite{Alkhateeb2019-v2}. Next we look at the mapping from the channel to the positions. Although the mapping from the channel to the positions may not always be unique, it is unique with a high probability in many practical wireless communication scenarios especially as the number of antennas increases which is widely exploited in the wireless fingerprinting \cite{fingerprint} and positioning \cite{positioning}. In other words, the mapping between positions and channel can be assumed bijective, so is the mapping between the uplink and downlink channels.

  \item In a TDD system, the channel reciprocity is usually assumed but in reality,  the analog radio front-ends at different wireless nodes such as  BS and the mobile users exhibit non-reciprocity due to non-identical behavior of the individual transmit and receive chains.  This is caused by the mismatches in the frequency-responses   of both the BS and user side radio front-ends between the transmit and receive modes, and  the differences in mutual coupling  of BS antenna units and the associated RF transceivers under transmit and receive modes  \cite{non-reciprocal} \cite{you}.

      Specifically, consider the channels between a BS with $N_t$ antennas and a single-antenna user in linear TDD systems. The uplink channel $\qh_u\in \mathbb{C}^{N_t \times1}$ and downlink channel $\qh_d\in \mathbb{C}^{N_t \times 1}$ can be written as \cite{reciprocity-tsp}\cite{Huang2019ULDL},
      \be
\underline{}        \qh_u =  \qR_u  \qh t_u , \mbox{~~~~~and ~~~~~}  \qh_d^T =  r_d  \qh^T  \qT_d,
      \ee
      where $\qh\in \mathbb{C}^{N_t \times1}$ is   the  physical reciprocal channel, $t_u$ and $r_d$ are the frequency-responses at the user side in the transmitting and receiving modes, respectively. Denote $\qL\in \mathbb{C}^{N_t \times N_t}$ as the frequency-response matrix   and $\qM \in \mathbb{C}^{N_t \times N_t}$ as the mutual coupling matrix of the BS, and then  $\qR_u = \qL_r \qM_r$ and  $\qT_d = \qM_t\qL_t$   where  the subscripts $t$ and $r$ denote the transmitting and receiving modes, respectively. The frequency-response matrix $\qL$ is diagonal but the mutual coupling matrix $\qM$ is not diagonal. In general, $\qM_r\ne \qM_t, \qL_r\ne \qL_t, t_u\ne r_d$, so the uplink and dowlink channels are non-reciprocal, and their relation can be described as
      \be
        \qh_d^T = r_d (\qR_u^{-1}\qh_u t_u^{-1} )^T \qT_{d}=r_d t_u^{-1}  (\qh_u  )^T (\qR_u^{-1})^T\qT_{d},
      \ee
      which is a deterministic and unique mapping.
\end{itemize}

Once the  bijective mapping between the uplink and downlink channel is established, it can be learned by using the deep neural networks based on   the universal approximation theorem \cite{universal}. Note that in the above, we have adopted explicit parametric modelling of uplink and downlink channels with simplifying assumptions (e.g., not all hardware impairment sources such as  power amplifier distortion, phase noise  and quantization noise, are considered), but in practice such parametric models may not be accurate. In this paper  we do not assume any specific parametric channel models in our theoretical development and instead we use the model-driven learning approach to learn the best mapping of the uplink to downlink channel in order to maximize the end performance.

 \section{The Proposed General Algorithm Framework}\label{algorithm}
 Since we consider a generic system which could be either TDD or FDD, the exact theoretical characterization on the mapping between the uplink and downlink channel is unknown as long as it is bijective, so the most viable way to obtain the downlink CSI without user feedback is to learn it from   data first, based on which, the downlink beamforming will be learned or optimized subsequently to maximize the sum rate.
  This is a traditional method that treats the channel learning and the end performance optimization separately. The main drawback of the separate learning is that the explicit channel learning process does not take into account the ultimate objective of maximizing the sum rate, and it   also causes error propagation when optimizing the beamforming. In this paper, we   use a deep learning approach to solve the problem \eqref{prob} directly from the uplink channel information. We   still include learning the downlink channel from the uplink channel information but it is only an intermediate step and the focus is not to achieve the best channel learning performance. The key idea of our proposed method is to exploit the optimal structure of the beamforming solution as the useful model information, which then guides the design of a highly efficient neural network to solve \eqref{prob},  with the assistance of the learned downlink CSI. More details are given below.

\subsection{Structure of beamforming}
 According to \cite{bjornson2014optimal},  the optimal downlink beamforming vectors that maximizes the sum rate  possesses the structure below
\begin{equation}\label{solution struc of sumrate}
  \qw_k^{\ast}=\sqrt{p_k}\frac{(\qI_{N_t}+\sum^K_{j=1}{\frac{q_k}{N_0}\qh_{D,j}\qh_{D,j}^H})^{-1}\qh_{D,k}}{\|(\qI_{N_t}+\sum^K_{j=1}{\frac{q_k}{N_0}\qh_{D,j}\qh_{D,j}^H})^{-1}\qh_{D,k}\|}, \forall k,
\end{equation}
where $p_k$ and $ q_k$ are positive parameters and satisfy $\sum_{k=1}^K p_k=\sum_{k=1}^K q_k=P$.  The parameter vector $\qp=[p_1,\ldots,p_K]^T$ represents the downlink power allocation.   $\qq=[q_1,\ldots,q_K]^T$ is an auxiliary vector variable which is useful to determine the direction of beamforming. Because it needs to satisfy the same total power constraint as the downlink power vector $\qp$,  $\qq$ can be interpreted as the virtual uplink power vector. The advantage of this representation is that the power vector   $[\qp; \qq]$ can be regarded as the key feature of the beamforming solution. Instead of learning the high-dimensional beamforming matrix $\qW$ directly, \eqref{solution struc of sumrate} allows us to learn the low-dimensional feature $[\qp; \qq]$, which will greatly improve the learning efficiency and accuracy, and reduce the training complexity.

\subsection{Key modules of the proposed neural network}
\begin{figure}[h]
\centering
 	\includegraphics[width=3.5in]{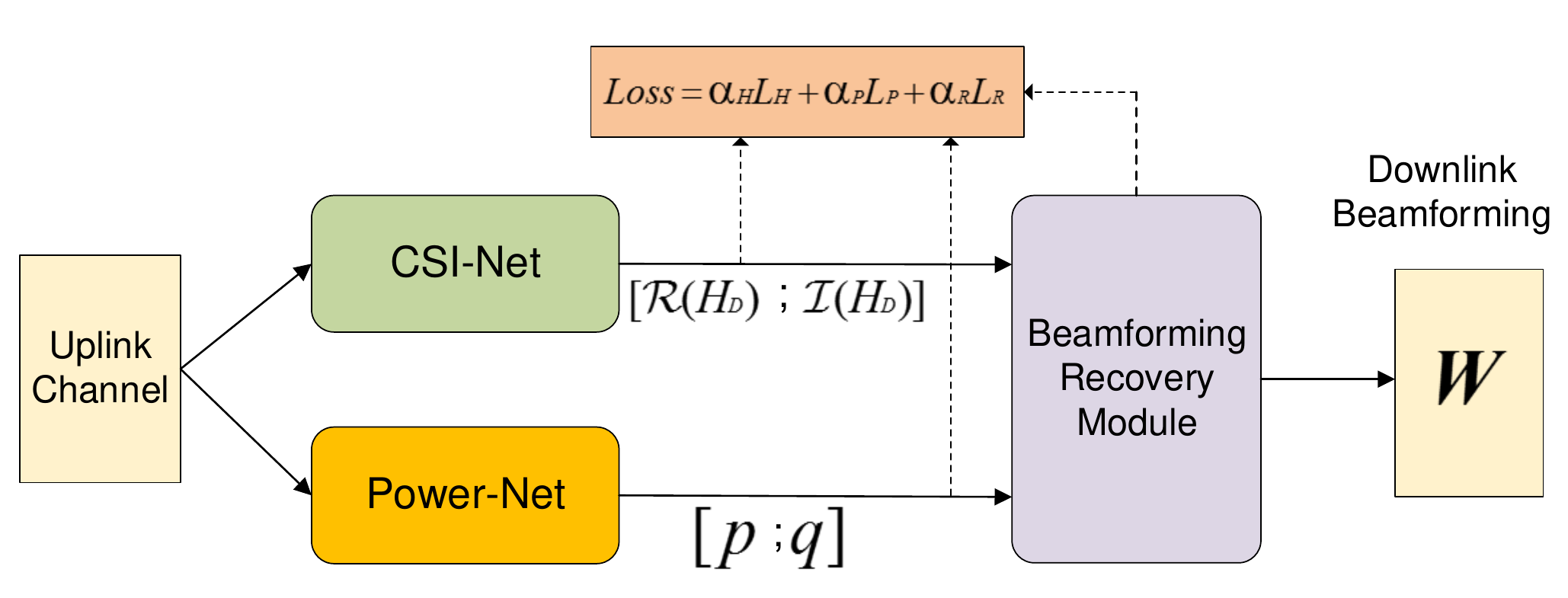}
 \caption{The proposed model-driven neural network structure.}
 \label{fig:NN}
\end{figure}

Based on the expression in \eqref{solution struc of sumrate}, we propose a neural network to jointly learn the downlink channel, the power feature vector with the end objective of maximizing the system sum rate, as illustrated in Fig \ref{fig:NN}.
The proposed neural network takes uplink channel information as input, and its output is the beamforming matrix $\qW$. The input to the neural network can be either  the perfect uplink CSI $\qH_U$ with a dimension of  $2N_t K$ by stacking the real and imaginary parts, or  the uplink pilot signal which will be discussed later in this section. The proposed neural network  structure consists of the following three modules:
\begin{itemize}
  \item \textbf{\textit{CSI-Net}}.  This sub-net aims to learn the downlink channel $\qH_D$ from the uplink channel and its output is separated into the real part $\mathcal{R}(\qH_D)$ and the imaginary part $\mathcal{I}(\qH_D)$. When the input is the uplink CSI, it will perform uplink-downlink calibration for TDD systems, while in FDD systems it will map the channel in the uplink band to the channel in the downlink band. It is not necessary to specify the TDD or FDD system as this module   learns the downlink channel automatically for either case. When the input is the uplink pilot signal, it   additionally estimates or refines the uplink channel but this is embedded in the module implicitly.   Suppose we have a channel training dataset in which there are $T$ uplink-downlink CSI pairs. Given the predicted output result of the $t$-th sample in the \textit{CSI-Net} is $\hat{\qH}_D^{(t)}$ and the target result in the training dataset is $\qH_D^{(t)}$, the  mean squared error (MSE)-based loss function of \textit{CSI-Net} is defined as
        \begin{equation}
        L_H=\frac{1}{2 T N_t K}\sum_{t=1}^T{||\qH_D^{(t)}-\hat{\qH}_D^{(t)}||^2}.
        \end{equation}
         The  structure of \textit{CSI-Net}   depends on the specific systems of interest, and in this paper, we adopt fully connected layers and details will be given in Section \ref{simu}. Note that for the uplink CSI input, when the channels for different users are uncorrelated and the  statistics is similar, we can use the same \textit{CSI-Net}  to learn the downlink channel in a single-user manner. This effectively increases the amount of training data by a factor of $K$, while reduces the complexity of the neural networks.

  \item \textbf{\textit{Power-Net}}. This sub-net aims to learn the concatenated power  vector $[\qp; \qq]$, which is the key feature of the beamforming solution.  In the literature,  there is no method available that can find the optimal $p^{\ast}_k$ and $q^{\ast}_k$ in \eqref{solution struc of sumrate} to maximize the sum rate with a reasonable complexity. The WMMSE algorithm \cite{shi2011iteratively} is a well known iterative method to find the locally optimal solutions. It   ensures the continuity of the mapping from the channel to the solution, which can be learned by a neural network. Therefore, we  generate samples  of the power allocation vectors $\qp$ and $\qq$ for training, by using the WMMSE algorithm. The supervised learning with the following loss function based on the MSE metric will be   used to train the \textit{Power-Net},
\begin{equation}\label{loss1}
  L_P=\frac{1}{2TK}\sum_{t=1}^T{\left(\|{\bf p}^{(t)}-\hat{\bf p}^{(t)}\|^2+\|{{\bf q}}^{(t)}-\hat{\bf q}^{(t)}\|^2\right)},
\end{equation}
where ${\bf p}^{(t)}$ and ${\bf q}^{(t)}$ are the $t$-th training samples of the downlink and  uplink power vectors in the power training database obtained from the WMMSE algorithm, respectively, and $\hat{\bf p}^{(t)}$ and $\hat{\bf q}^{(t)}$ are the predicted results of   \textit{Power-Net}. Similar to \textit{CSI-Net}, the neural network structure is system-dependent, and in this paper, we adopt   fully-connected layers   or convolutional neural network (CNN) layers which will be specified in simulation results of Section \ref{simu}.

  \item \textbf{\textit{Beamforming Recovery Module.}}  This module has two functions. First, it aims to find the downlink beamforming matrix from the downlink channel output from the \textit{Power-Net} and the uplink and downlink power output from the \textit{Power-Net} using the structure specified in \eqref{solution struc of sumrate}; there is no parameter to optimize in this module. Second, this module is important to calculate the sum rate using unsupervised learning which   then  forms the overall loss function for effective hybrid training as described in the next subsection.
\end{itemize}

\subsection{Hybrid training}
To train the proposed neural network, we   construct the overall loss function as the weighted sum of the losses of the \textit{CSI-Net}, the \textit{Power-Net} and the sum rate, i.e.,
\be\label{loss}
    \mbox{Loss} = \alpha_H L_H + \alpha_P L_P + \alpha_R L_R,
\ee
where $L_R = -R^{\text{sum}}$, and $\alpha_i, i \in\{H, P, R\}$ is the weight for each loss component. $L_H$ and $L_P$ can be obtained by supervised learning from  the \textit{CSI-Net} and the \textit{Power-Net}, respectively,  while $R^{\text{sum}}$ can be calculated by using the beamforming matrix obtained from the above \textit{Beamforming Recovery Module} and the learned downlink channel from the \textit{CSI-Net} based on \eqref{sinr};   overall, the training adopts a hybrid supervised and unsupervised approach.
Note that both channel learning and power learning are  auxiliary in our proposed algorithm, and the focus of the overall learning is to maximize the  sum rate but not to achieve the best learning performance of the downlink channel matrices and power vectors. The incorporation of the sum rate into the loss function is important because this ensures that the training of the neural network is guided by the end performance; this is in stark contrast to the separate training approach which only focuses on learning the channel or the power vectors but cannot guarantee  the sum rate performance. In addition, the inclusion of the $L_H$ and $L_P$ is also important because the downlink channel is unknown and it is difficult to learn the overall mapping from the uplink channel directly to the downlink beamforming which   leads to unsatisfactory training performance.

\begin{figure}[t]
\centering
 	\includegraphics[width=2.5in]{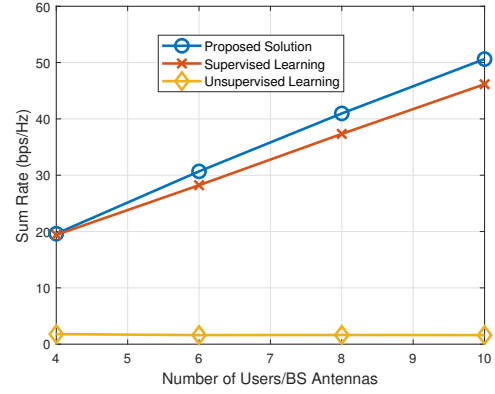}
 \caption{Comparison of the sum rate performance of the proposed algorithm, supervised learning and unsupervised learning methods versus the number of users/antennas ($K=N_t$), $P = 20$ dB.}
 \label{fig_rate_vs_K_super_unsuper}
\end{figure}

We illustrate the advantage of the proposed algorithm over the supervised learning ($\alpha_R=0$) and unsupervised learning methods ($\alpha_H=\alpha_P=0$) in Fig. \ref{fig_rate_vs_K_super_unsuper}, where the uplink CSI follows distribution and the relation between the $i$-th  elements of the uplink and downlink CSI is $h_d[i] =(h_u[i])^2$.
As can be seen, the proposed algorithm outperforms the supervised learning method, and the performance gap increases as the number of antennas grows. When $N_t=K=10$, the proposed algorithm achieves about 10\% higher sum rate than the supervised learning method, while the unsupervised learning cannot achieve satisfactory sum rate performance as explained above.  The results show that in sharp contrast to the case where perfect CSI is available, existing methods of supervised and unsupervised learning   could not achieve satisfactory performance when the CSI   needs to be learned.

\subsection{Learning from the uplink pilot signal}
In this subsection, we introduce the preprocessing when the uplink channel information is in the form of the received pilot signal. Assume that $K$ users transmit $L$ pilot symbols. Suppose the pilot signals sent by all users  are collected in $\qX\in \mathbb{C}^{K\times L}$, then  the received pilot signal in the uplink $\qY\in \mathbb{C}^{N_t\times L}$ can be written as
\be
    \qY = \qH_U\qX + \qN,
\ee
where $\qN\in \mathbb{C}^{N_t\times L}$ is the received noise and the elements of $\qN$ has zero mean and variance of $\sigma^2_n$. In order to recover the uplink channel from the pilot signal, we choose the pilot $\qX$ to be a sub-matrix of  scaled   discrete Fourier transform (DFT) matrix of dimension $\max(K,L)\times \max(K,L)$ (multiplied by the square root of the transmit power), which has orthogonal columns and rows and its elements have unit amplitude.

For our proposed algorithm, instead of using the received pilot $\qY$, we use the least square version $\tilde \qY\in \mathbb{C}^{N_t\times K}$ as the input to the neural network:
\begin{equation}
\tilde \qY\triangleq \frac{\qY\qX^H}{LP}  =
\begin{cases}
\qH_U + \frac{\qN \qX^H}{LP} & L\ge K,\\
\qH_U\frac{\qX\qX^H}{LP} + \frac{\qN \qX^H}{LP}   & L<K.
\end{cases}
\end{equation}
Clearly when $L\ge K$, $\qX\qX^H=LP\qI$ which means in this case, pilot sequences between users are orthogonal; while when $L<K$, there exists pilot contamination among users which will degrade the performance of estimating the uplink channel. Note that in our proposed approach, we do not need to estimate the uplink channel explicitly, but we learn the downlink beamforming from the uplink pilot directly.

For comparison, the traditional separate approach would first estimate the uplink channel, e.g., by using a linear minimum mean-squared error (MMSE) estimator, and then use it as the input of the neural network. Suppose the  linear MMSE estimator is in the form of
\be\label{HU}
    \hat \qH_U =  \qY \qR + \qB,
\ee
where $\qR$ and $\qB$ are the weighting coefficients. Then the corresponding channel estimation can be obtained by solving the following optimization problem:
\be\label{mse}
    \min_{\qR, \qB} \|\qY\qR+\qB - \qH_U\|^2.
\ee
The optimal $\qR$ and $\qB$ are given by
\bea\label{mmse_coeff}
    \qR^* &=&   \left(\qX\qQ\qX^H+ N_t \sigma_n^2\qI\right)^{-1}\qX^H\qQ,\\
    \qB^* &=& -\bar\qH_U (\qX\qR-\qI),
\eea
and the liner MMSE estimation of $\qH_U$ is given by
\be
    \hat\qH_U =  \qY \qR + \qB =
    \bar \qH_U + (\qY - \bar \qH_U\qX) \left(\qX\qQ\qX^H+ N_t \sigma_n^2\qI\right)^{-1}\qX^H\qQ,
\ee
where $\bar\qH_U$ is the mean value of $\qH_U$, i.e., $\qH_U = \bar \qH_U + \Delta \qH_U$, $\Delta\qH_U$ has zero mean, and $\qQ\triangleq E[\Delta \qH_U^H \Delta \qH_U]$ which can be estimated from the training data of the uplink channel. The derivation can be found in Appendix A.

\section{Low-complexity Implementation for   Massive MIMO Downlink}\label{massive}
The proposed algorithm framework in Section \ref{algorithm} is general so it can be applied to massive MIMO downlink, straightforwardly. However, the large number of antennas may introduce a high computational complexity,  especially in the calculation of the loss function during the training. In this section, we propose two techniques that can reduce the complexity when calculating the loss function of the sum rate in  the training process for massive MIMO systems,   without compromising the end performance, especially when $N_t\gg K$.

\subsection{Massive MIMO channel model}
We assume that the BS is equipped with  uniform linear array (ULA) antennas, without loss of generality.
The channel between the BS and a user (the user index is omitted for simplicity) that consists of $L_p$ paths is modelled as
\be\label{channel:massive}
\qh= \sum\limits_{l = 1}^{L_p} \alpha_l e^{ - j2\pi f\tau_l + j\phi_l} \qa(\theta_l),  
\ee
where $\alpha_l, \tau_l, \phi_l$ and $\theta_l$ are the attenuation, the path delay, the phase shift and the angle of arrival (AoA, for the uplink) or the angle of departure (AoD, for the downlink)  of the $l$-th path, respectively. Moreover, $\qa(\theta)\in \mathbb{C}^{N_t\times 1}$ is the   array response vector defined as
\be
\qa (\theta)= \left[ {1,e^{ -j\frac{2\pi}{\lambda}d\sin(\theta)}, \cdots ,e^{- j\frac{2\pi}{\lambda} d (N_t - 1)\sin (\theta)}} \right]^T,
\ee
where $d$ is  the antenna spacing and $\lambda$ is the wavelength.

From the structure of the optimal beamforming \eqref{solution struc of sumrate}, we can see that it involves inversion of a matrix of dimension $N_t\times N_t$. This   causes a high complexity for massive MIMO systems, because   the calculation of the sum rate in the loss function \eqref{loss} requires the construction of the optimal beamforming according to \eqref{solution struc of sumrate}, which needs to calculate the inversion of a matrix of size $N_t\times N_t$.

\subsection{Dimension reduction}
In the first technique, we aim to reduce the dimension of the inverse matrix,  when calculating the optimal beamforming using \eqref{solution struc of sumrate}. We find the following proposition is useful  which is adopted from the result in interference channels \cite{chara_MISO} \cite{beam_hetero_MIMO}.

\emph{Proposition 1:} Suppose $N_t\ge K$ and the users' channels are linearly independent and that $\qh_{i}^H \qh_{j}\ne 0, \forall i\ne j$. Then if $\qw_k$   is a beamforming vector for user $k$ that corresponds to a rate point on the Pareto boundary, there exists complex numbers  $\{\xi_{ij}\}^K_{i,j=1}$ such that
\be
    \qw_k = \sum_{k=1}^K\xi_{ik} \qh_i, \sum_{k=1}^K \|\qw_k\|^2= P.
\ee
It can be proved using the same method as that in \cite{chara_MISO} and  therefore the proof is omitted.

Recall that $\qH_D=[\qh_{D,1}, \cdots, \qh_{D,K}]$, and define $\bm\xi_k=[\xi_{1k}, \cdots, \xi_{Kk} ]^T \in \mathbb{C}^{K\times 1} $. Suppose $\qw_k = \qH  \bm\xi_k$, then the SINR expression becomes
\be
\gamma_k=\frac{|\mathbf{h}_{D,k}^H \qH_D  \bm\xi_k|^2}{\sum_{j\neq k}^K|\mathbf{h}_{D,k}^H \qH_D  \bm\xi_j|^2+N_0}.
\ee

Define the eigenvalue decomposition $\qH_D^H\qH_D= \qU \Lambda \qU^H$, where $ \qU \in \mathbb{C}^{K\times K}$ is the unitary eigen-matrix and $\Lambda \in \mathbb{C}^{K\times K}$ is the diagonal eigenvalue matrix. Now define the new beamforming vector $\qv_k=\Lambda^{1/2}\qU^H_D\bm\xi_k$ and the new channel vector $\qg_k = \Lambda^{-1/2}\qU^H\qH_D^H\qh_{D,k}$. Then the sum rate maximization can be written equivalently as
\bea\label{prob_reduced}
\max_{\qv_1,\cdots, \qv_K} && \sum_{k=1}^K \log_2(1+\gamma_k) \\
\gamma_k&=&\frac{|\qg_k^H  \qv_k|^2}{\sum_{j=1, j\neq k}^K|\qg_k^H \qv_j|^2+N_0},\notag\\
&& \sum_{k=1}^K \|\qv_k\|^2\le P. \notag
\eea
We can see from the above new problem formulation \eqref{prob_reduced} that the size of the new channel matrix $\{\qg_k\}$ reduces from $N_t\times K$ to $K\times K$. The structure of the optimal beamforming is revised to
\begin{equation}\label{solution struc of sumrate_reduced}
  \qv_k^{\ast}=\sqrt{p_k}\frac{(\qI_K+\sum^K_{j=1}{\frac{q_k}{N_0}\qg_{j}\qg_{j}^H})^{-1}\qg_{k}}{\|(\qI_{K}+\sum^K_{j=1}{\frac{q_k}{N_0}\qg_{j}\qg_{j}^H})^{-1}\qg_{k}\|}, \forall k,
\end{equation}
and as a result the size of the matrix inversion is reduced from $N_t\times N_t$ to $K\times K$.

Note that although  the above dimension reduction technique reduces the dimension for matrix inversion significantly, it involves extra matrix multiplication and eigenvalue decomposition, when constructing the new channel vectors . It was shown in \cite{stable}  that standard linear algebra operations for a square matrix of dimension $n\times n$, including   matrix inversion and eigenvalue decomposition problems have the same time complexity as the matrix multiplication algorithm and there exists the Coppersmith-Winograd algorithm for matrix multiplication with a complexity of $\mathcal{O}(n^{2.376})$ \cite{algorithm}. Therefore,  the benefit of the dimension reduction technique  on the overall training time is only obvious when $N_t\gg K$ and this is verified by the simulation results in Section \ref{simu}.

\subsection{ZF beamforming in the loss function}
Another technique to reduce the complexity of matrix inversion in \eqref{solution struc of sumrate}, when calculating the sum rate in the loss function, is to use the ZF beamforming in the following form
\be\label{BFZF}
    \qW = d\qH_D (\qH_D^H\qH_D)^{-1},
\ee
where $d$ is chosen to satisfy the power constraint, i.e., $\|\qW\|^2=P$. It has been shown in the seminal work on massive MIMO \cite{Scaling} that, ZF beamforming is near-optimal when the number of antennas is large. From the complexity's viewpoint, the advantage of the ZF beamforming is that it only involves the matrix inversion of $\qH_D^H\qH_D$ which is a $K\times K$ matrix and reduces the complexity of matrix inversion in \eqref{solution struc of sumrate}.

Similar to the dimension reduction technique, the advantage of the ZF beamforming is more prominent when $N_t\gg K$ and diminishes as $K$ increases, but its performance is always close to the WMMSE solution as long as $N_t\gg 1$. These properties are  verified by the simulation results in Section \ref{simu}.

\section{Multicell Massive MIMO Downlink}\label{multicell}
\begin{figure}[!htbp]
\centering
\includegraphics[width=0.3\textwidth, trim={0cm 0.0cm 0cm 0.0cm},clip]{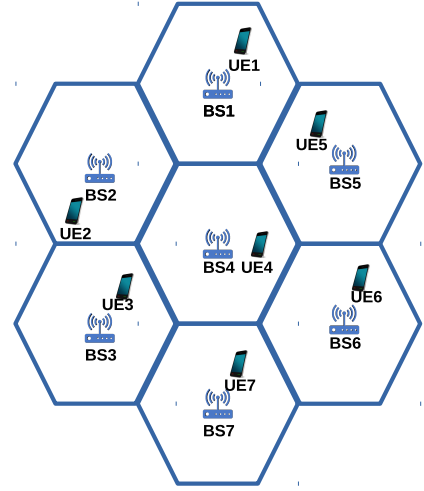}
\caption{An multicell massive MIMO network illustration with 7 cells.}
\label{fig system setup multicell}
\end{figure}

In this section, we apply the proposed model-based learning to  multicell massive MIMO systems, where an example of the multicell massive MIMO system with seven cells is illustrated in Fig. \ref{fig system setup multicell}. Specifically, we   first introduce  the distributed multicell massive MIMO downlink beamforming, without the need of signal or data exchange between cells, and then describe the uplink channel estimation via pilots, followed by the proposed model-based learning method.
\subsection{Distributed Optimization of  Beamforming in Multicell Massive MIMO Downlink}
Consider an $N_c$-cell massive MIMO system, and in each cell   a  BS with $N_t$ antennas serves $K_i$ single-antenna users, $i=1,\dots, N_c$.
Denote the beamforming vector for the $k_j$-th user in the $j$-th cell as $\mathbf{w}_{k_j} \in \mathbb{C}^{N_t \times 1}$, then the received scalar signal $y_{k_i}$ in the $i$-th cell is expressed as follows:
\bea
y_{k_i} &=& {\mathbf{h}^H_{D, k_i, i} \mathbf{w}_{k_i} {x}_{k_i}} + \underbrace{\sum_{k_j=1, k_j \neq k_i}^{K_i} \mathbf{h}^H_{D,k_i, i} \mathbf{w}_{k_j} {x}_{k_j}}_{\text{Intra-cell Interference}}\notag \\
 &&+ \underbrace{\sum_{j=1, j\neq i}^{N_c} \sum_{k_j=1}^{K_j} \mathbf{h}^H_{D, k_i, j} \mathbf{w}_{k_j} {x}_{k_j}}_{\text{Inter-cell Interference}} + n_{k_i},
\eea
where $\mathbf{h}_{D, k_i, j} \in \mathbb{C}^{N_t \times 1}$ is the downlink channel from the BS in the $j$-th cell to the $k_i$-th user in the $i$-th cell, ${x}_{k_j} $ is the signal for the $k_j$-th user in the $j$-th cell. Therefore the SINR   of the $k_i$-th user in the $i$-th cell is written as follows
\begin{equation}
\label{eq SINR multicell}
\gamma_{\text{SINR},k_i} = \frac{|\mathbf{h}^H_{D,k_i, i} \mathbf{w}_{k_i}|^2}{\sum_{k_j=1, k_j \neq k_i}^{K_i} |\mathbf{h}^H_{D,k_i, i} \mathbf{w}_{k_j}|^2 + \sum_{j=1, j\neq i}^{N_c} \sum_{k_j=1}^{K_j} |\mathbf{h}^H_{D,k_i, j} \mathbf{w}_{k_j}|^2 + N_0}.
\end{equation}
Based on \eqref{eq SINR multicell}, to calculate  $\gamma_{\text{SINR},k_i}$ for the $k_i$-th user in the $i$-th cell, it requires not only the beamforming $\mathbf{w}_{k_i}$ and downlink channels $\mathbf{h}_{D,k_i, i}$, but also   the knowledge of any possible interfering cell $j$ with $j=1,\dots, N_c$ and $j\neq i$. This includes interfering BSs' downlink channels to the considered user $k_i$ in the $i$-th cell $\mathbf{h}_{D, k_i, j}$  and all users' beamforming vectors in those interfering cells $\mathbf{w}_{k_j}$ with $k_j = 1,\dots, K_j$. This means to optimize the SINR in any single cell, the solution of the beamforming vector in this cell is coupled with the solutions in other cells. Traditional methods, such as the coordinated transmission,  treat the multicell as a large single cell and use ZF beamforming method to cancel out the interference term in \eqref{eq SINR multicell}. These methods require the cooperation between cells such as the exchange of CSI and centralized joint optimization of beamforming vectors, which   results in high signaling overhead \cite{CIT-069}.

In order to decouple the beamforming between different cells, the signal to leakage plus noise ratio (SLNR) is used in this paper as an alternative, where the SLNR   of the $k_i$-th user in the $i$-th cell is defined as follows \cite{sadek2007leakage}:
\begin{equation}
\label{eq SLNR multicell}
\gamma_{\text{SLNR}, k_i} =  \frac{|\mathbf{h}^H_{D,k_i, i} \mathbf{w}_{k_i}|^2}{\sum_{j=1}^{N_c} \sum_{k_j=1, k_j \neq k_i}^{K_j} |\mathbf{h}^H_{D, k_j, i} \mathbf{w}_{k_i}|^2 + N_0}.
\end{equation}
Note that the key difference between the SLNR in \eqref{eq SLNR multicell} and the SINR in \eqref{eq SINR multicell} is the interference item in the denominator, where the SLNR in \eqref{eq SLNR multicell} considers the `leaked' interference power due to the beamforming $\mathbf{w}_{k_i}$ from the $k_i$-th user in the $i$-th cell to all other cells' users, instead of the received interference in SINR in \eqref{eq SINR multicell}.
 Traditionally the data rate is defined based on SINR in \eqref{eq SINR multicell}, but this requires the cooperation between different cells to estimate the data rate performance of the user. Since the SLNR definition in \eqref{eq SLNR multicell} shares  similarity with the SINR definition in \eqref{eq SINR multicell}, here we define an approximate data rate based on SLNR the $k_i$-th user in the $i$-th cell as follows:
\begin{equation}\label{eq R SLNR multicell}
R_{\text{SLNR}, k_i} = \log_2(1+\gamma_{\text{SLNR}, k_i}) \approx \log_2(\gamma_{\text{SLNR}, k_i}),
\end{equation}
where the subscript ``SLNR'' is used to differentiate it from the original data rate definition, and the approximation is based on the high SNR assumption.

With the definition of SLNR in \eqref{eq SLNR multicell},   the optimization of beamforming vector $\mathbf{w}_{k_i}$ for the $i$-th cell depends only on the downlink channels from the BS in the $i$-th channel to users in all cells, which can be estimated via the uplink estimation. More importantly, since the required channel information can be obtained by the BS in a single cell, the exploitation of SLNR instead of SINR decouples the beamforming in individual cells. Therefore the  multicell sum rate maximization problem  can be decoupled to the per-cell sum rate maximization problem to be addressed in each cell as
\begin{equation}
\label{eq overall sum rate multicell}
\max_{\mathbf{w}_{k_i}}   \sum_{k_i=1}^{K_i} R_{\text{SLNR}, k_i}, \quad \text{s.t. }   \sum_{k_i=1}^{K_i} \|\mathbf{w}_{k_i}\|^2 \leq P, i\in \{1,\dots, N_c\},
\end{equation}
where $P$ is the transmit power limit for each BS. To solve the problem in  \eqref{eq overall sum rate multicell}, the beamforming vector $\mathbf{w}_{k_i} $ is rewritten as $\mathbf{w}_{k_i} = \sqrt{p_{k_i}} \mathbf{u}_{k_i}$, where $\mathbf{u}_{k_i}$ satisfies $||\mathbf{u}_{k_i}|| = 1$ for all $k_i = 1, \dots, K_i$, and represents the direction of the beamforming vector, while the power is characterized by $p_{k_i}$.
Then with \eqref{eq R SLNR multicell}, the objective function in \eqref{eq overall sum rate multicell} can be further rewritten as follows:
\bea
\label{eq percell objective rewritten multicell}
&&\sum_{k_i=1}^{K_i} R^{\text{SLNR}}_{k_i} \approx \sum_{k_i=1}^{K_i} \log_2\left(\gamma_{\text{SLNR}, k_i}\right) \\
&&= \log_2 \left(\prod_{k_i=1}^{K_i}\frac{p_{k_i}|\mathbf{h}^H_{D,k_i, i} \mathbf{u}_{k_i}|^2}{\sum_{j=1}^{N_c} \sum_{k_j=1, k_j \neq k_i}^{K_j} p_{k_i} |\mathbf{h}^H_{D, k_j, i} \mathbf{u}_{k_i}|^2 + N_0}\right),\notag
\eea
while the per-cell sum rate maximization problem in \eqref{eq overall sum rate multicell} can be further rewritten as follows
\begin{equation}
\begin{aligned}
\label{eq percell sum rate fin multicell}
&\min_{p_{k_i}, \mathbf{u}_{k_i}} & \prod_{k_i=1}^{K_i} \left( \frac{N_0}{|\mathbf{h}^H_{D,k_i, i} \mathbf{u}_{k_i}|^2} {p_{k_i}}^{-1} + \frac{\sum_{j=1}^{N_c} \sum_{k_j=1, k_j \neq k_i}^{K_j} |\mathbf{h}^H_{D, k_j, i} \mathbf{u}_{k_i}|^2 }{|\mathbf{h}^H_{D,k_i, i} \mathbf{u}_{k_i}|^2} \right)\\
&\text{ s.t.} & \sum_{k_i=1}^{K_i} p_{k_i}  \leq P.
\end{aligned}
\end{equation}
The problem \eqref{eq percell sum rate fin multicell} is non-convex, so we propose to use the alternating optimization to solve it. Specifically, with a given power allocation $p_{k_i}$, the optimal beamforming direction $\mathbf{u}_{k_i}$  is solved as:
\begin{equation}
\label{eq direction multicell}
\mathbf{u}_{k_i}^{*} = \frac{ \left(\mathbf{I}_{N_t} + \frac{p_{k_i}}{N_0} \sum_{j=1}^{N_c} \sum_{k_j=1}^{K_j} \mathbf{h}_{D, k_j, i} \mathbf{h}^H_{D, k_j, i}\right)^{-1} \mathbf{h}_{D, k_i, i}}{||  \left(\mathbf{I}_{N_t} + \frac{p_{k_i}}{N_0} \sum_{j=1}^{N_c} \sum_{k_j=1}^{K_j} \mathbf{h}_{D, k_j, i} \mathbf{h}^H_{D, k_j, i}\right)^{-1} \mathbf{h}_{D, k_i, i} ||}, \forall k_i,
\end{equation}
When the beamforming directions $\mathbf{u}_{k_i}$ are fixed, the objective function of \eqref{eq percell sum rate fin multicell} is a posynomial with regard to $p_{k_i}$, so \eqref{eq percell sum rate fin multicell}  can be solved via geometric programming \cite{chiang2007power}. Since in each step, the objective function is nondecreasing, such an alternating optimization algorithm will converge. Therefore we can use the alternating optimization algorithm to generate the downlink power solutions as the labelled data in the training process.

When $N_t > \sum_{j=1}^{N_c} K_j$, the dimension reduction techniques in Section V.B can be also applied to the multicell scenario to reduce the training complexity.

Note that the optimal beamforming based on SLNR involves only the downlink channels from the BS in a single cell to the users in all cells. This enables the distributed beamforming   at each single cell and no cooperation is required between cells, which helps to reduce the signaling overhead  in the multicell scenario. More importantly, this  enables the proposed learning-based method to be  decoupled in the multicell scenario and applicable when the system scales up.

\subsection{Learning From Uplink Pilots}
The optimization based on SLNR requires the channel from the BS in each cell to learn the downlink channels to  users from the uplink channel information, which can be achieved by estimation from uplink pilots.

Consider the $i$-th cell surrounded by $N_c-1$ neighbouring cells. Assume that each user  in the multicell   system transmits pilot symbols of length $L$, then for the BS in the $i$-th cell, the received pilot signal in the uplink $\mathbf{Y}_i \in \mathbb{C}^{N_t \times L}$ is written as
\begin{equation}
\mathbf{Y}_i = \mathbf{H}_{U,i} \mathbf{X} + \mathbf{N}_i,
\end{equation}
where $\mathbf{H}_{U,i} \in \mathbb{C}^{N_t \times \sum_{j=1}^{N_c} K_j}$ denotes the uplink channels from all users to the BS in the $i$-th cell, $\mathbf{N}_i \in \mathbb{C}^{N_t \times L}$ is the received noise whose elements has zero mean and variance of $\sigma_i^2$, and $\mathbf{X} \in \mathbb{C}^{\sum_{j=1}^{N_c} K_j\times L}$ is the pilots sent by all users. Similar to the single cell scenario, the pilot $\mathbf{X}$ adopts the sub-matrix of a discrete DFT matrix with dimension $\max\{\sum_{j=1}^{N_c} K_j, L\} \times \max\{\sum_{j=1}^{N_c} K_j , L\}$, which is scaled by the square root of the transmit power. Then the least square version of the uplink channel $\mathbf{\tilde{Y}}_i \in \mathbb{C}^{N_t \times \sum_{j=1}^{N_c} K_j}$ is used as the input of the neural network, and can be estimated via the received pilot signal $\mathbf{Y}_i$ as follows
\begin{equation}
\label{eq ytilde multicell}
\mathbf{\tilde{Y}}_i = \frac{\mathbf{Y}_i \mathbf{X}^H}{LP} = \left\lbrace\begin{matrix}
\mathbf{H}_{U,i} + \frac{\mathbf{N}_i \mathbf{X}^H}{{LP}}, L\geq \sum_{j=1}^{N_c} K_j,\\
\mathbf{H}_{U,i} \frac{\mathbf{X}\mathbf{X}^H}{LP}  + \frac{\mathbf{N}_i \mathbf{X}^H}{{LP}}, L<\sum_{j=1}^{N_c} K_j.
\end{matrix}\right.
\end{equation}
Clearly to avoid the pilot contamination between users, it requires the pilot's length $L$ to be no less than the total number of users, i.e., $L \geq \sum_{j=1}^{N_c} K_j$.

For the traditional separate approach, the uplink channel will be estimated using the linear MMSE method.

\subsection{The Proposed Distributed Learning}
In this subsection, we will adapt the proposed model-based learning approach in Section \ref{algorithm} to the multicell scenario. We will use a different SLNR beamforming structure below based on \eqref{eq direction multicell}, i.e.,
\begin{equation}
\label{eq direction multicell2}
\mathbf{w}_{k_i}^{*} = \sqrt{p_{k_i}}\frac{ \left(\mathbf{I}_{N_t} + \frac{p_{k_i}}{N_0} \sum_{j=1}^{N_c} \sum_{k_j=1}^{K_j} \mathbf{h}_{D, k_j, i} \mathbf{h}^H_{D, k_j, i}\right)^{-1} \mathbf{h}_{D, k_i, i}}{||  \left(\mathbf{I}_{N_t} + \frac{p_{k_i}}{N_0} \sum_{j=1}^{N_c} \sum_{k_j=1}^{K_j} \mathbf{h}_{D, k_j, i} \mathbf{h}^H_{D, k_j, i}\right)^{-1} \mathbf{h}_{D, k_i, i} ||}, \forall k_i.
\end{equation}

From \eqref{eq direction multicell2}, we can see that in order to construct the multicell beamforming, we need the downlink channel information and downlink power allocation. Therefore, we can still use the hybrid loss function \eqref{loss} which is rewritten below
\be\label{loss_MC}
    \mbox{Loss} = \alpha_H L_H + \alpha_P L_P + \alpha_R L_R,
\ee
except the loss of  {\textit{Power-Net}} only involves the downlink power, i.e., for the $i$-th cell, it becomes
\begin{equation}\label{loss1}
  L_P=\frac{1}{2TK_i}\sum_{t=1}^T\sum_{i=1}^{K_i}{\left(\|{p_{k_i}}^{(t)}-\hat{p}_{k_i}^{(t)}\|^2 \right)}.
\end{equation}

Note that the proposed method based on SLNR and learning from uplink pilots can be generalized to larger systems. The above analysis is based on the $i$-th cell, which treats all other cells as interfering cells. With a homogeneous assumption that the conditions are similar in each cell in the whole multicell massive MIMO system, the analysis of $i$-th cell is applicable to any cell in the system. This means the trained neural network's parameters are applicable to all cells  in a distributed manner, which is demonstrated in the simulation results in Section \ref{simu}.C.

\section{Simulation Results}\label{simu}
In this section, numerical simulations are carried out to evaluate the performance of the proposed  algorithms.  $10^5$ training samples and $10^3$ testing samples are used with  a  batchsize of 100 and 200   epochs.   We use Keras with Tensorflow 1.15 as the backend to train the proposed neural network on a  Nvidia GP100 GPU card in a High Performance Computing (HPC) system.  The weight factors used in the loss function are hyper-parameters and chosen as $\alpha_H=\alpha_P=1, \alpha_R = 0.001$, by using a trial and error approach, unless otherwise specified. The much lower value of $\alpha_R$ is to balance each component in the loss function considering typical values of the sum rate.

For the proposed model-driven solution, when calculating the sum rate in the loss function,  both the original algorithm that uses the beamforming solution in \eqref{solution struc of sumrate}, and the one that uses the simplified ZF beamforming in \eqref{BFZF} are included. For comparison, we consider the following benchmark algorithms:
\begin{itemize}
\item The WMMSE  solution:   this is the solution  obtained by using the iterative algorithm proposed in \cite{shi2011iteratively} assuming the downlink channel is available, therefore it serves as a performance upper bound.
\item The Learned Channel and Beamforming Solution: this solution is obtained by first learning the downlink channel via supervised learning, then using the learned downlink channel to infer the beamforming solution using unsupervised training.
 \item The Learned Channel and ZF Beamforming: this solution is used as the benchmark solution in the single-cell scenario. It   first learns the downlink channel via supervised learning as the above solution, and then constructs a ZF beamforming \eqref{BFZF} by using the learned channel.
\item The Learned Channel and  SLNR Beamforming: this solution is used as the benchmark solution in the multi-cell scenario. It first learns the downlink channel via supervised learning as the above two solutions, and then constructs a   SLNR beamforming \eqref{eq direction multicell} by using the learned channel. Note that the SLNR based approximate data rate is used only for the purpose of the neural network training, while the sum rate performance is calculated based on the actual SINR defined in \eqref{eq SINR multicell}.
\end{itemize}
 Note that the closed-form non-iterative ZF and SLNR based beamforming solutions are chosen as the benchmarks to ensure the low complexity  comparable to our proposed algorithm.  In the following, we will present the simulation results and analysis for three scenarios: single-cell small-scale MIMO, single-cell  massive MIMO, and multicell massive MIMO systems, as well as the generalization results.

\subsection{Small-scale MIMO Scenario}
In this scenario, a TDD downlink system in which one BS with equal numbers of transmit antennas and users is considered, i.e., $N_t=K\ngg 1$.
 We assume the uplink channel elements follow an independent  Rayleigh distribution with zero mean and unit variance. Following the result in \cite{Huang2019ULDL},   the relation between the downlink channel and the uplink channel due to the radio  front-end mismatch is modelled as:
\be
    \qh_D= c\Phi\qh_U,
\ee
where  the  unitary matrix $\Phi$ and $c\sim\mathcal{CN}(0,1)$ are used to model the  mismatches in the frequency-responses of  the  BS and the user sides, respectively. This mapping will be learned by the \textit{CSI-Net}.  Suppose the learned downlink channel is $\hat \qH_D$. The learning performance is characterized by the normalized MSE (NMSE), which is defined as
\be
    \mbox{NMSE} = E\left[\frac{\|\hat \qH_D - \qH_D\|^2}{\|\qH_D\|^2}\right].
\ee
The structure and hyper-parameters of the proposed neural network are as follows. For the \textit{CSI-Net},  we use  four fully connected layers, each with $4KN_t$  neurons and  the  `tanh' activation function. No activation function is employed at the output layer.  The \textit{Power-Net} employs four fully connected layers,  each with  $4KN_t$ neurons and the `relu' activation function and batch normalization. The output layer of the \textit{Power-Net}   uses `softmax' as the  activation function. Beamforming learning for the `Learned Channel and Beamforming' solution  uses the same neural network  as \textit{CSI-Net}, except that a batch normalization is included at each layer.  $10^6$ training samples are used for this scenario. We also provide brief analysis of the complexity for the online prediction. For fully connected layers, suppose the input dimension is $M$ and the number of neurons in the hidden layer is $N$, then the numbers of multiplication and addition operations  are  equal to $MN$. Therefore the overall neural network has an approximate complexity $\mathcal{O}\left(N_t^2K^2\right)$ for the online prediction.

\begin{figure}
\centering
\subfigure[]
{
	\begin{minipage}{3.1in}
    \label{fig2:a}
    \centering
	\includegraphics[width=2.5in]{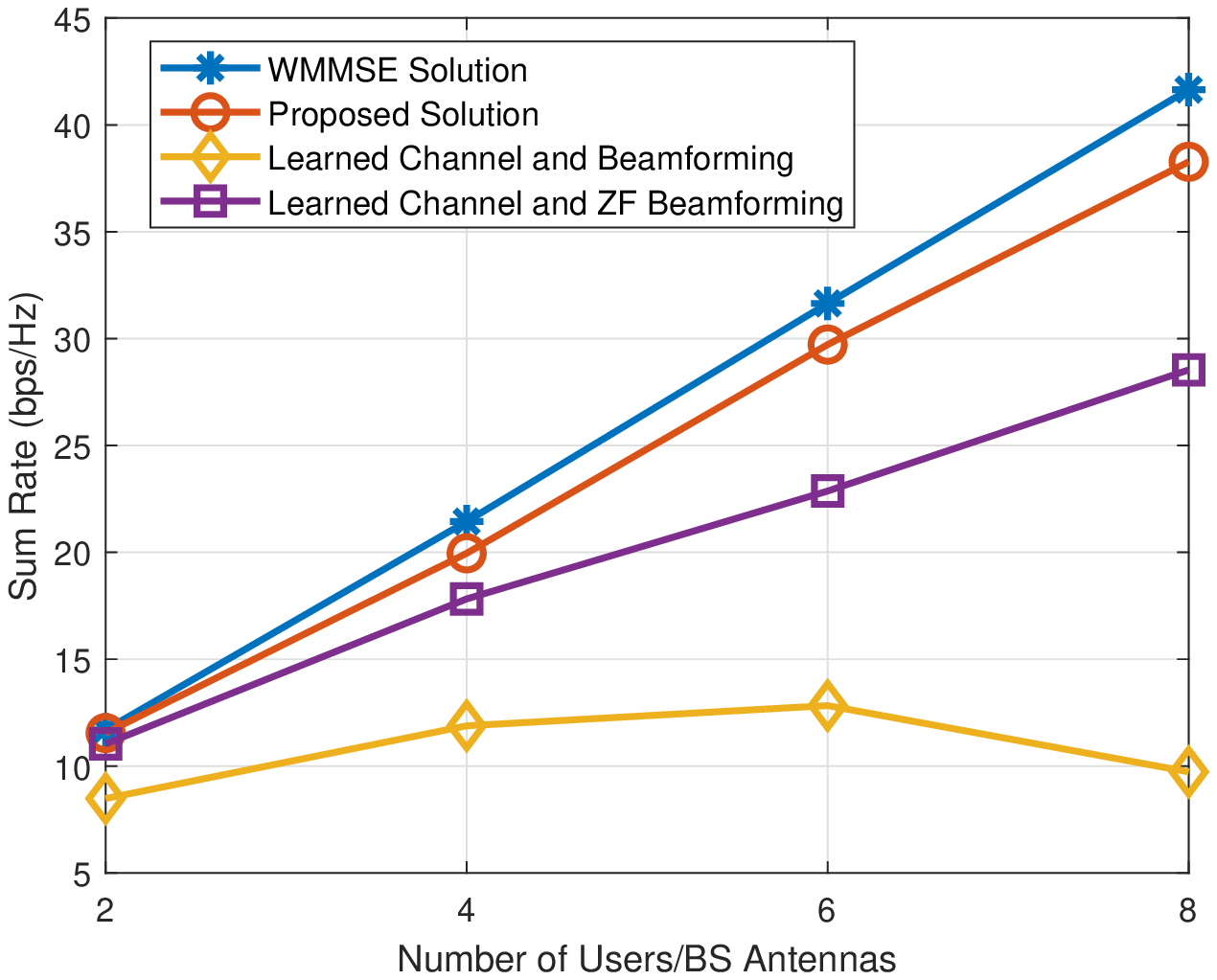}  \label{fig_rate_vs_K}
	\end{minipage}
}
\subfigure[]
{
	\begin{minipage}{3.1in}
    \label{fig2:b}
    \centering
	\includegraphics[width=2.5in]{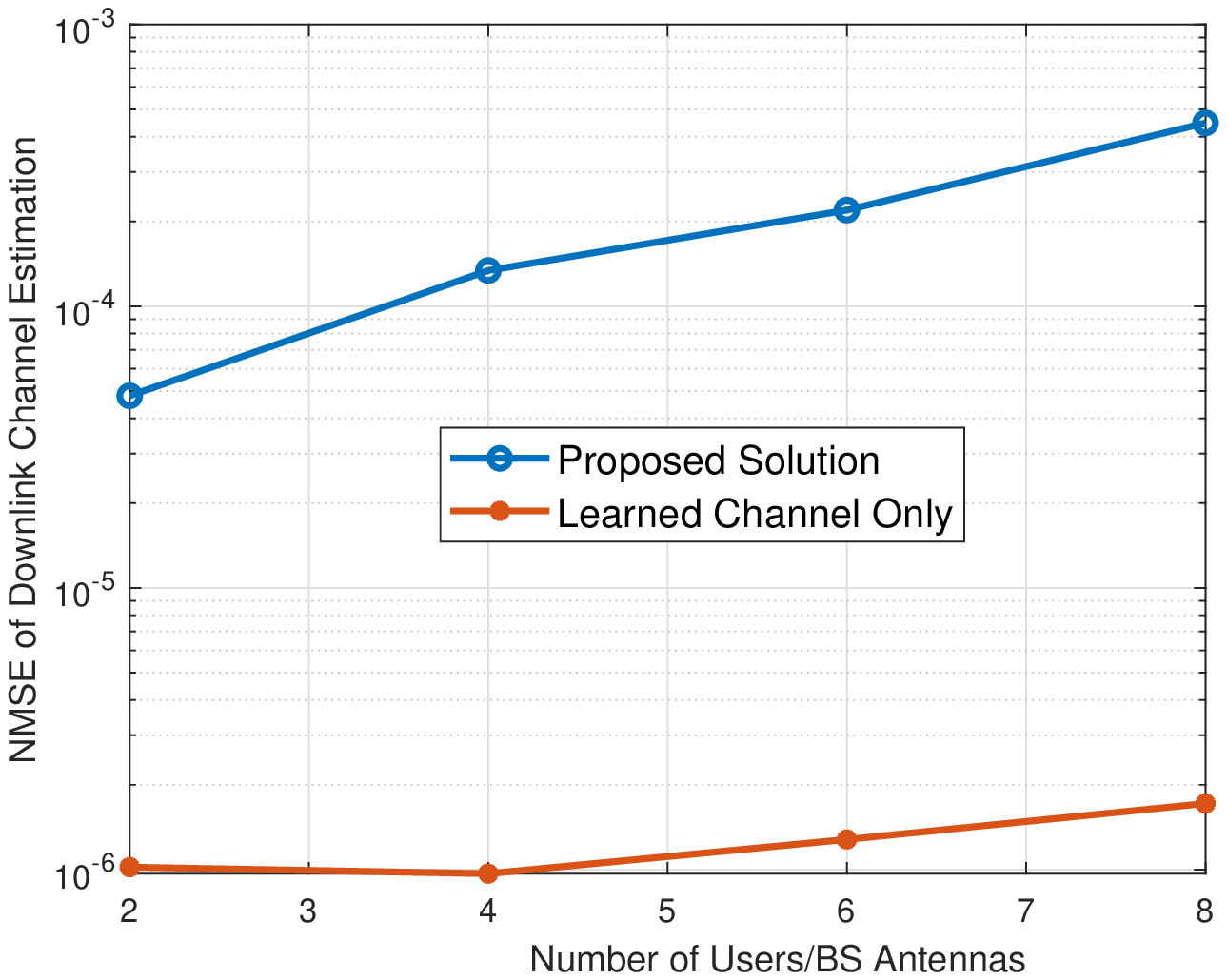}\label{CE_vs_K}
	\end{minipage}
}
\centering
\caption{(a) Comparison of the sum rate versus the number of users/antennas, $P = 20$ dB; (b) Comparison of the channel estimation performance versus the number of users.}
\end{figure}

We first show the sum rate results of various algorithms in Fig. \ref{fig_rate_vs_K}, when the transmit power is 20 dB and the number of users/BS antennas vary from 2 to 8. We assume that the perfect uplink channel CSI $\qH_U$ is available.  It can be seen that the proposed solution achieves the sum rate   close to that of the WMMSE algorithm , and it significantly outperforms the benchmark schemes.  The performance of both benchmark methods that first learn the downlink channel is not satisfactory. The scheme of learned channel and beamforming, achieves the worst performance while the one using a ZF beamforming achieves a higher data rate but still much lower than the proposed solutions. To further investigate the reason, we plot the channel learning results of downlink channel learning in Fig. \ref{CE_vs_K}. It is obvious that the benchmark schemes that explicitly learns the channel first outperforms the proposed solution in terms of achieving a lower estimation NMSE. However, its end performance, i.e., the sum rate is much lower  than the proposed solution. This confirms the advantage of the proposed model-driven learning with a hybrid training that can better exploit the available uplink channel information to optimize the end performance;   the traditional approach that separately learns the channel and then designs the beamforming is not adequate since it is purely data-driven.

\subsection{Single-cell Massive MIMO Scenario}
 In this scenario, we adopt the channel model specified in Section \ref{massive} for  a FDD massive MIMO system and assume $N_t=64$.  The uplink and downlink operate at 2.5 GHz and 2.4 GHz, respectively. The antenna spacing is half wavelength of the downlink signal. Since the DoA or the AoD $\theta$ is   limited within a certain region of the mean angle $\bar \theta$,   it is modelled as $\theta \in [\bar\theta-\Delta \theta/2,\bar\theta+\Delta \theta/2], -\frac{\pi}{6}\le \Delta \theta\le \frac{\pi}{6}$. The path delay $\tau$ and the phase shift $\phi$ are uniformly distributed in the ranges of $[0, 10^{-4}]$ and $[-\pi, -\pi]$, respectively. We assume the downlink channel attenuation $\alpha_{d,l}$ of   the $l$-th path follows an  independent  Rayleigh distribution with zero mean and unit variance.
 Given the ULA channel model in \eqref{channel:massive}, the nonlinear relation between the uplink channel $\qh_D$ and dowlink channel $\qh_D$ are characterized by:
 \bea\label{channel:massive:nonlinear}
\qh_U&=& \sum\limits_{l = 1}^{L_p} \alpha_{u,l} e^{ - j2\pi f\tau_l + j\phi_l} \qa_u(\theta_l),~~~~\notag \\
\qh_D&=& \sum\limits_{l = 1}^{L_p} c\Phi(\alpha_{u,l})^2 e^{ - j2\pi f\tau_l + j\phi_l} \qa_d(\theta_l),
\eea
where  the  unitary matrix $\Phi$ and $c\sim\mathcal{CN}(0,1)$ are used to model the  mismatches in the frequency-responses of  the  BS and the user sides, respectively, which will be learned by the \textit{CSI-Net}.

The structure and the hyper-parameters of the proposed neural network are as follows. The \textit{CSI-Net} contains three fully connected  layers, each with $2N_t K$ neurons each and the `tanh' activation function.   The  \textit{Power-Net} uses two 1-D CNN with filter sizes of 16 and 8, batch normalization and `relu' activation function and a dropout rate of 0.3, followed by a fully connected layer with 256 neurons and `relu' activation function, and an output layer that uses `softmax' as the  activation function.  We provide brief complexity analysis for the online prediction. For $L$ 1-D convolutional layers, suppose there are $c_l$ kernels of size $a_l$ in   the $l$-th convolutional layer and the input dimension is $b\times d_l$ (padding is added such that the first input and output dimensions remain the same across layers, i.e., $b$),  then the output dimension of  is $b\times c_l$, and the numbers of multiplication and addition operations  are   equal to  $a_l b c_{l-1} c_l$ when $l>1$ and $a_l b d_{l} c_l$ when $l=1$.  Thus, the total   complexity of all convolutional layers measured by the number of multiplications  is $\mathcal{O}\left(\sum_{l=1}^L a_l b   c_{l-1} c_l\right)$. Therefore for the \textit{Power-Net}, the complexity for the online prediction is approximately $\mathcal{O}\left(N_t K \right)$, while the overall complexity is still $\mathcal{O}\left(N_t^2 K^2 \right)$ considering the \textit{CSI-Net}.

\begin{figure}
\centering
\subfigure[]
{
	\begin{minipage}{3.1in}
    \label{fig2:a}
    \centering
	\includegraphics[width=2.5in]{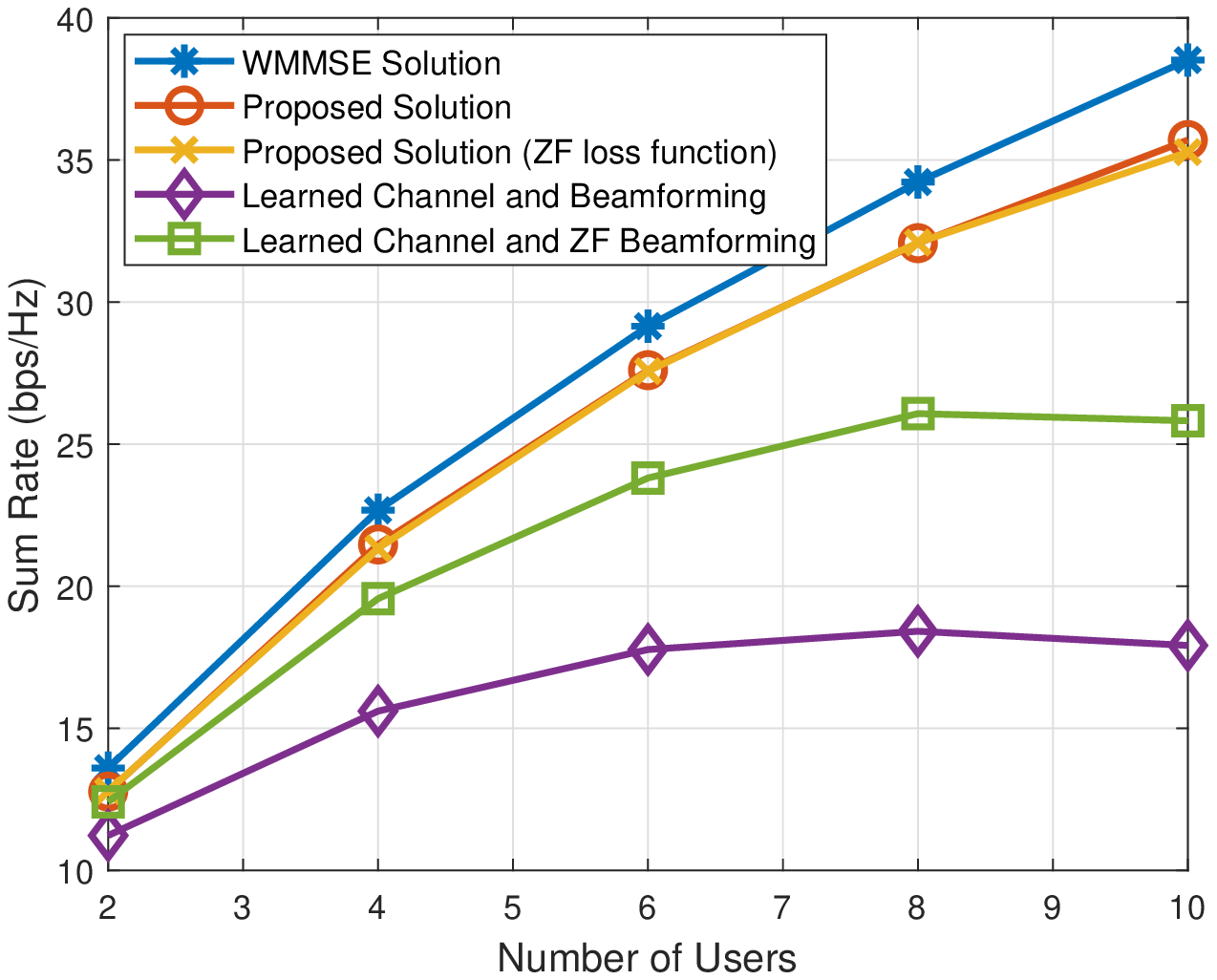}  \label{fig_rate_vs_K_massive}
	\end{minipage}
}
\subfigure[]
{
	\begin{minipage}{3.1in}
    \label{fig2:b}
    \centering
	\includegraphics[width=2.5in]{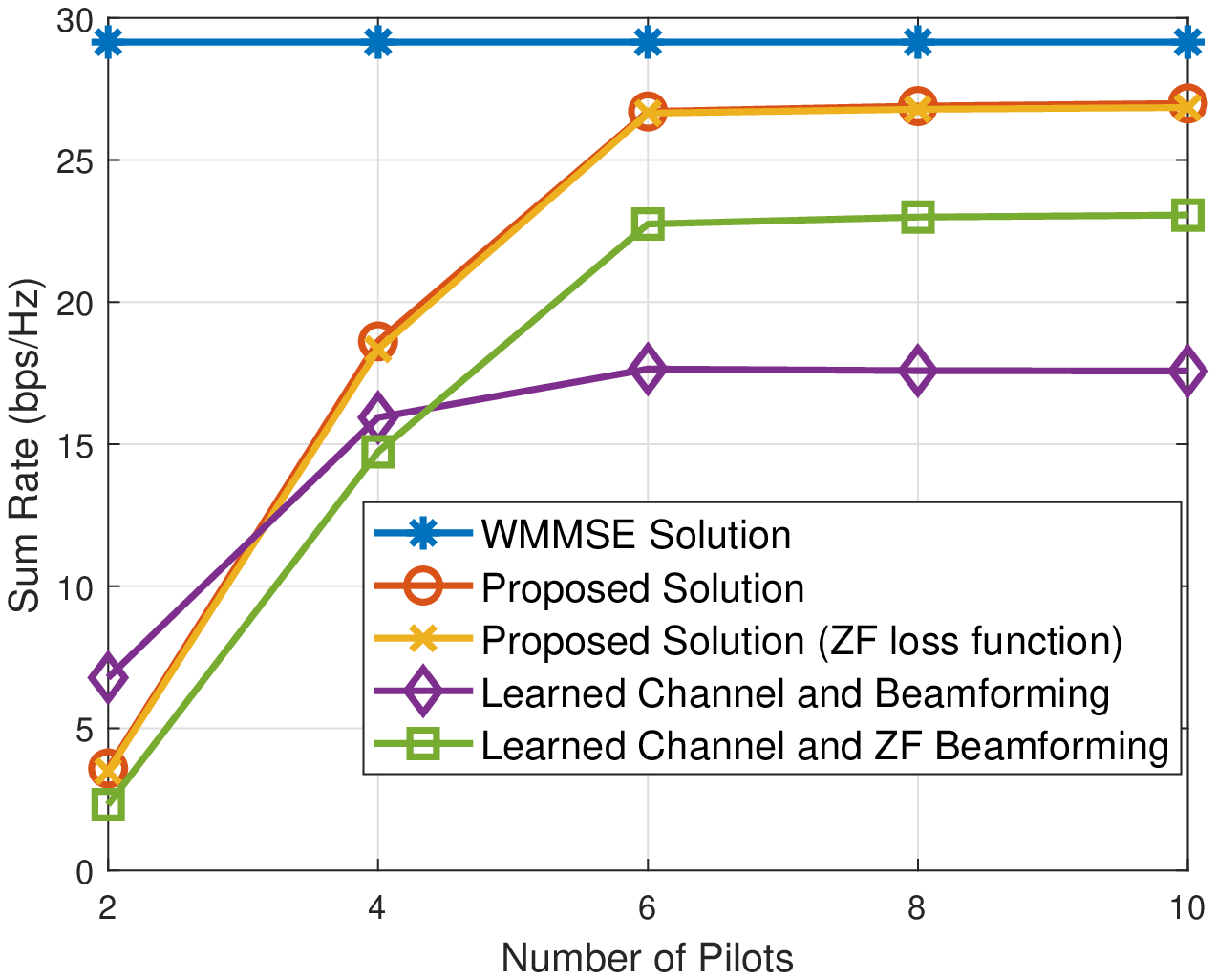}\label{fig_rate_vs_pilots_massive}
	\end{minipage}
}
\centering
\caption{Comparison of the sum rate versus the number of users, $N_t=32$, $P = 10$ dB, versus (a) the number of users and (b) the number of pilots in the single-cell massive MIMO scenario.}
\end{figure}

  The sum rate results of various algorithms are shown in Fig. \ref{fig_rate_vs_K_massive},  when the transmit power is 10 dB and the number of users vary from 2 to 10. It can be seen that the proposed solution achieves the sum rate very close to the WMMSE solution and the use of a ZF beamforming in the loss function has almost no performance loss. The proposed solutions outperform  the learned channel and the ZF beamforming solution, by about 10\%, although the latter performs much better than the small scale MIMO scenario. The learned channel and beamforming solution is still the worse because it uses a data-driven approach, without taking into account the overall design objective when learning the downlink channel.

  Next, we evaluate the required training time of the proposed algorithms when using the low-complexity implementation discussed in Section \ref{massive}. The training time of the proposed original algorithm $T_0$ and the percentage of the time required by the reduced dimension  and the ZF beamforming techniques in relation to $T_0$ are shown in Table \ref{run_time} at the top of the next page. It is obvious that when the number of users $K$ is small, the reduced dimension in matrix inversion of the two low-complexity schemes leads to a much shorter training time. However, as the number of users $K$ increases, the gain of the low-complexity schemes diminishes. This is because for the reduced dimension scheme, it involves an extra eigenvalue decomposition and matrix multiplications. In addition, as the number of users $K$ increases, the training time is dominated by the width of the fully connected layers which is $2N_t K$, therefore the time saved by matrix inversion becomes insignificant.

 The sum rate performance versus the number of pilots is shown in Fig. \ref{fig_rate_vs_pilots_massive}, when the number of users is $K=6$. Similar to Fig. \ref{fig_rate_vs_K_massive}, it is confirmed again that the use of ZF beamforming in the loss function achieves almost the same performance as the original algorithm and both are close to the WMMSE solution when the number of pilot symbols  $L\ge 6$. The learned channel and ZF beamforming achieves good performance in this scenario, which is different from the small scale scenario. There is still significant gap between the proposed solutions and the learned channel and beamforming solution and this verifies the superior performance of the proposed model-driven approach.

\begin{table*}[!htbp]
\begin{center}
\begin{tabular}{ |c|c|c|c| c| c|}
 \hline
  \diagbox[dir=NW]{Proposed Algorithms}{No. of Users} & $K=2$ &$K=4$  & $K=6$&$K=8$ & $K=10$\\  \hline
  Original formulation, $T_0$ (s) & 1570 & 2314    & 3388 & 4716 & 6683 \\  \hline
  Reduced dimension (\%) &  67.24\%  &  80.98\%  &  102.86\%  &  95.68\%  &   99.98\%\\  \hline
  ZF Loss (\%) & 60.11\%   & 73.69\%   & 84.02\% &   87.78\%  &   93.13\%\\  \hline
\end{tabular}\vspace{5mm}
\caption{Training time of the proposed algorithms for massive MIMO systems, $N_t=64$.}
\label{run_time}
\end{center}
\end{table*}

\subsection{Multicell Massive MIMO Scenario}
In the multicell massive MIMO scenario, we consider a TDD system in which the uplink and the downlink operate at different frequencies of 2.5 GHz and 2.4 GHz, respectively.  In the simulations, a multicell system with $N_c = 7$ cells is considered, where each cell has one active user as illustrated in Fig. \ref{fig system setup multicell}. The total transmit power of the BS is 10 dBm, the bandwidth is 20 MHz and the noise power spectrum density is -174 dBm/Hz. The radius of each cell is 200 m, and the users are randomly located in each cell by following a uniform distribution. The minimum distance between each user and its serving BS is 10 m.

 The channel attenuation includes both the small and large scale fading effects. The relation between small scale channel attentions is the same as \eqref{channel:massive:nonlinear} in the single-cell scenario.  The large scale fading $\alpha_{\text{PL}}^{(dB)}$ (measured in dB, and  the uplink/downlink subscript is omitted) between the BS and the user is given as follows
\begin{equation}
\alpha_{\text{PL}}^{(dB)} = p_0 + \beta\log_{10}(d),
\end{equation}
where $p_0$ is the reference path loss gain measured in dB, which includes the effect of the central frequency $f_c$, $\beta$ is related to the path loss exponent, and $d$ is the distance between the BS and the user measured in kilometer. In the simulation, we assume that the uplink and downlink path loss gains are expressed as, respectively,
\begin{equation}
\alpha_{\text{PL,u}}^{(dB)} = 127 + 30\log_{10}(d),\mbox{~and~} \alpha_{\text{PL,d}}^{(dB)} = 128.1 + 37.6\log_{10}(d).
\end{equation}

The structure and the hyper-parameters of the proposed neural network are as follows.   Three fully connected layers are used in \emph{CSI-Net} and each has $2   N_t   N_c^2$ neurons each and use the `tanh' activation function. The weight factors used in the loss function are $\alpha_H=1, \alpha_R=-1$.
During the training procedure, this \emph{CSI-Net} is reused for each cell to learn the downlink channels based on their uplink channels. Since the constructed \emph{CSI-Net} and the proposed method are for a single cell which do not rely on signalling exchange with other cells, they can be  deployed in a distributed way at each BS once the training is completed.  The complexity for the online prediction is approximately $\mathcal{O}\left(N_t^2 N_c^2 \right)$.

\begin{figure}
\centering
\subfigure[]
{
	\begin{minipage}{3.1in}
    \label{fig2:a}
    \centering
	\includegraphics[width=2.5in]{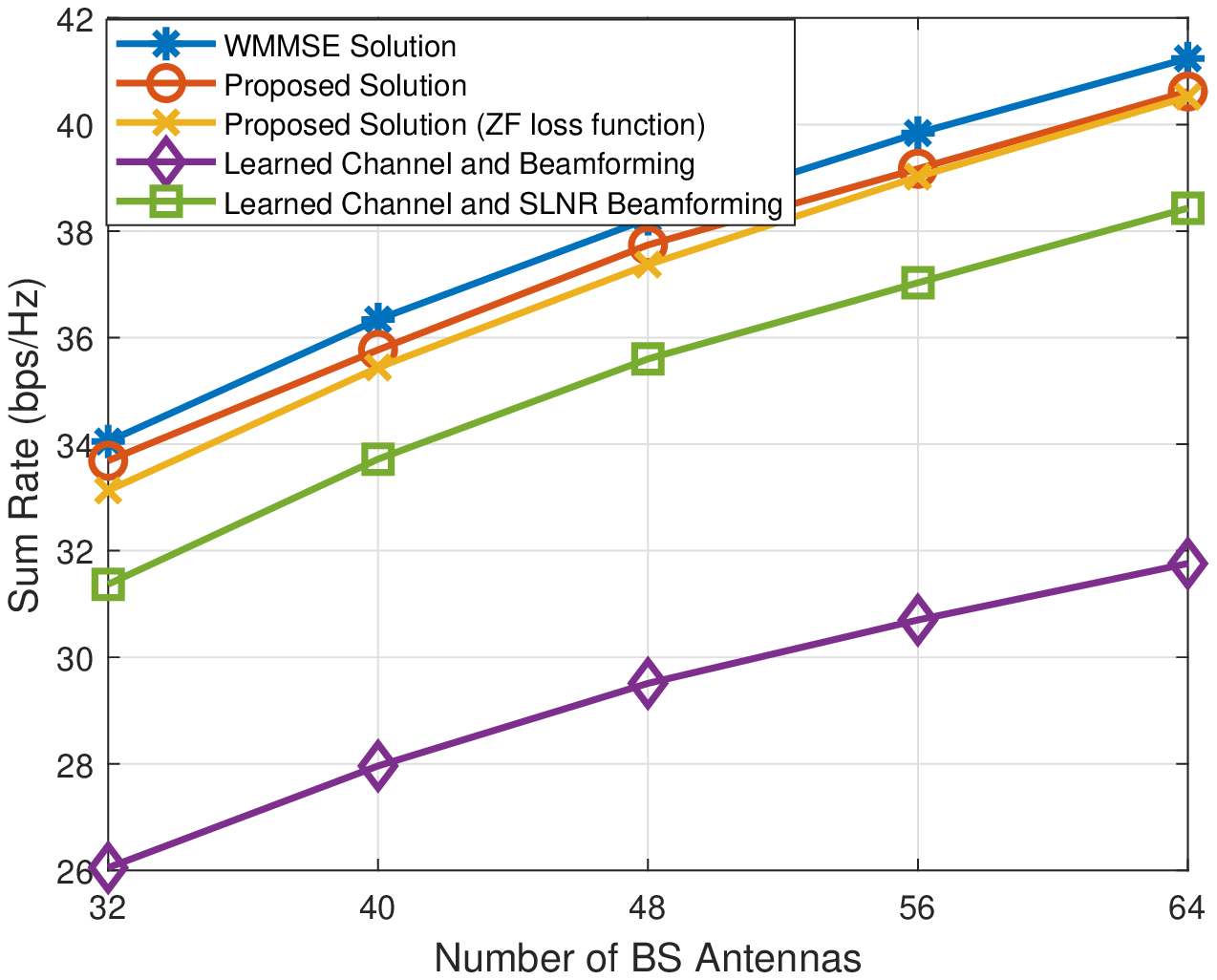}  \label{fig sumrate vs Nt multicell}
	\end{minipage}
}
\subfigure[]
{
	\begin{minipage}{3.1in}
    \label{fig2:b}
    \centering
	\includegraphics[width=2.5in]{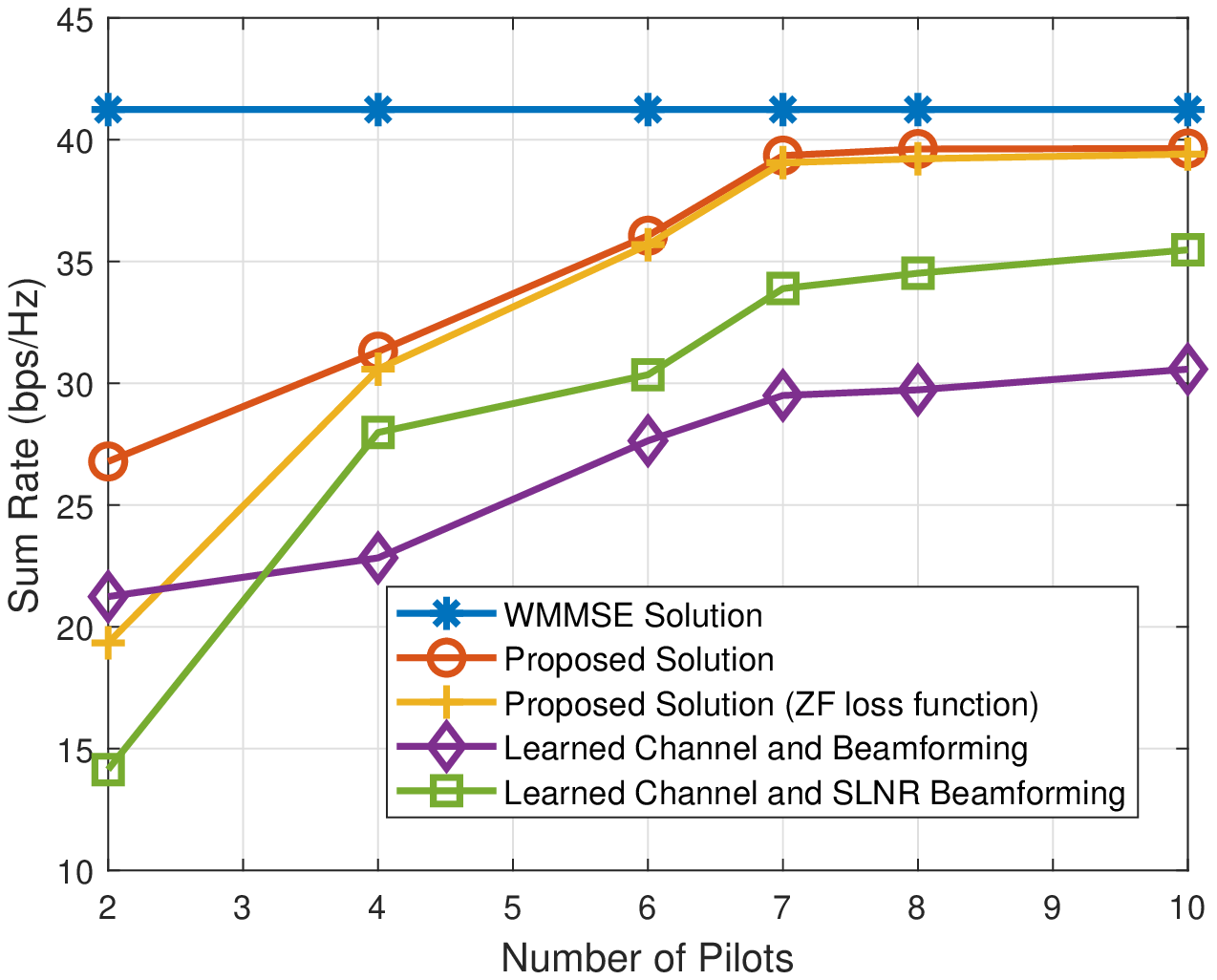}\label{fig sumrate vs L multicell}
	\end{minipage}
}
\centering
\caption{Comparison of the sum rate in the multicell scenario versus (a) the number of BS antennas,  $P = 10$ dBm and (b) the number of pilots in the multicell scenario, $N_t=64$, $P = 10$ dBm.}
\end{figure}

In the multicell massive MIMO scenarios, the number of transmit antennas is one of the key factors that determines the users' end performance, so we first evaluate the sum rate performance of all users   versus different numbers of antennas  in Fig. \ref{fig sumrate vs Nt multicell}. As seen from Fig. \ref{fig sumrate vs Nt multicell}, the proposed solution achieves a tight sum rate performance compared to the WMMSE solution. The proposed solution outperforms all benchmark algorithms, while the performance comparison of each benchmark algorithm follows a similar trend as in the single-cell scenario. It is also noticed that the proposed solution based on the ZF loss function  shows a close performance to the proposed solution based on the SLNR in the loss function, and the performance gap is generally reducing as    the number of antennas  increases. It is also clear that the data-driven approaches are worse than the model-driven approaches, while the learned channel and beamforming solution is still the worst.

 Next we investigate the sum rate performance versus different numbers of pilot symbols $L$ in Fig. \ref{fig sumrate vs L multicell}, when  $N_t=64$ and the uplink channels are estimated via the pilots. It can be seen that the performance of all algorithms improves as the number of pilots increases, but there is a significant gap from the WMMSE solution when $L<7$.   When $L\ge 7$, the proposed solutions with SLNR and ZF loss functions achieve close performance to the WMMSE solution and significantly outperform other benchmark schemes. This observation is expected as there are a total of seven users in the considered scenario. When the number of pilots is less than seven, there will be users with non-orthogonal pilots,  which results in a deteriorated accuracy of the recovered uplink channel from the pilots.

\subsection{Generalization}
\begin{figure}
\centering
\subfigure[]
{
	\begin{minipage}{3.1in}
    \label{fig2:a}
    \centering
	\includegraphics[width=2.5in]{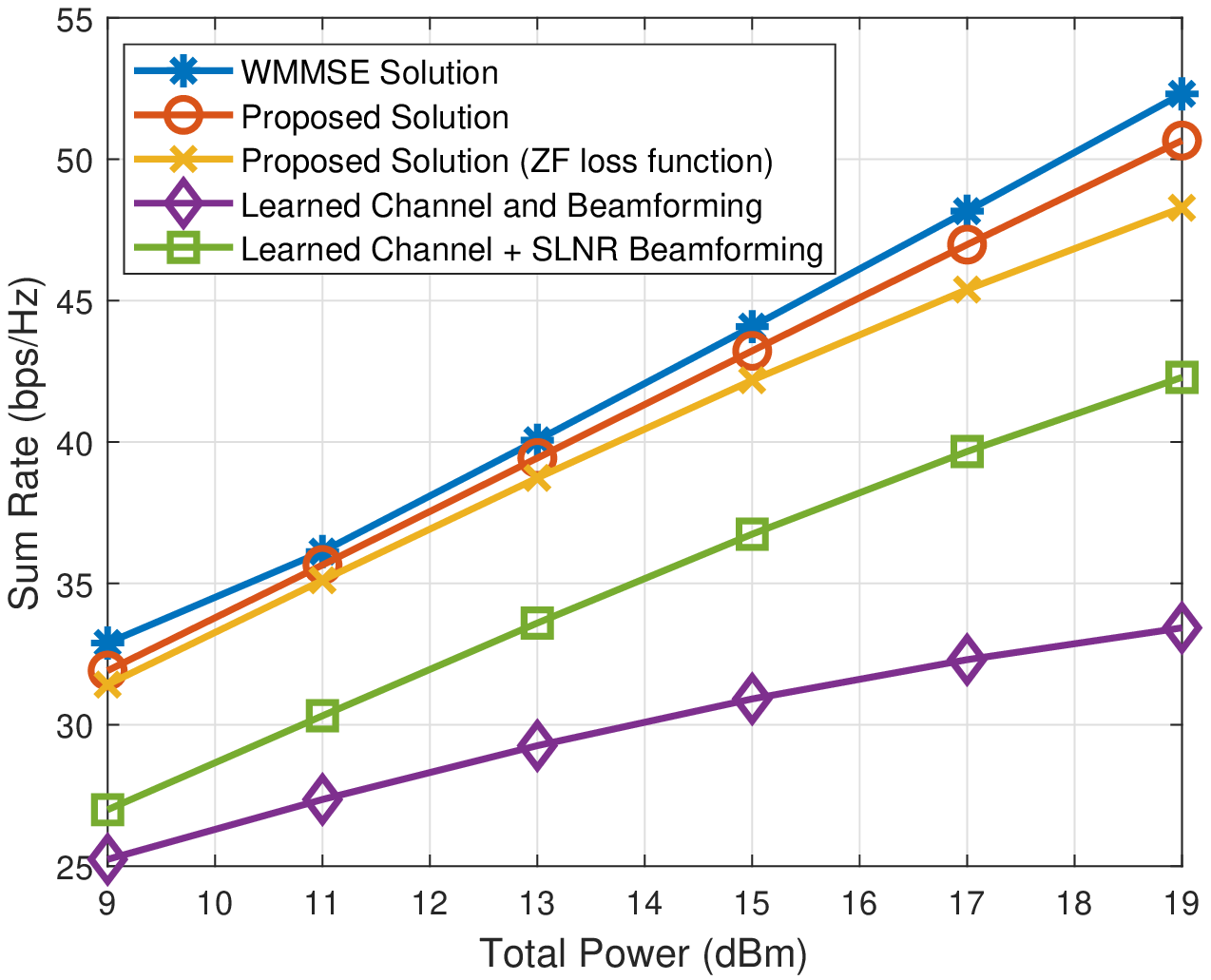}  \label{fig PdBm multicell}
	\end{minipage}
}
\subfigure[]
{
	\begin{minipage}{3.1in}
    \label{fig2:b}
    \centering
	\includegraphics[width=2.5in]{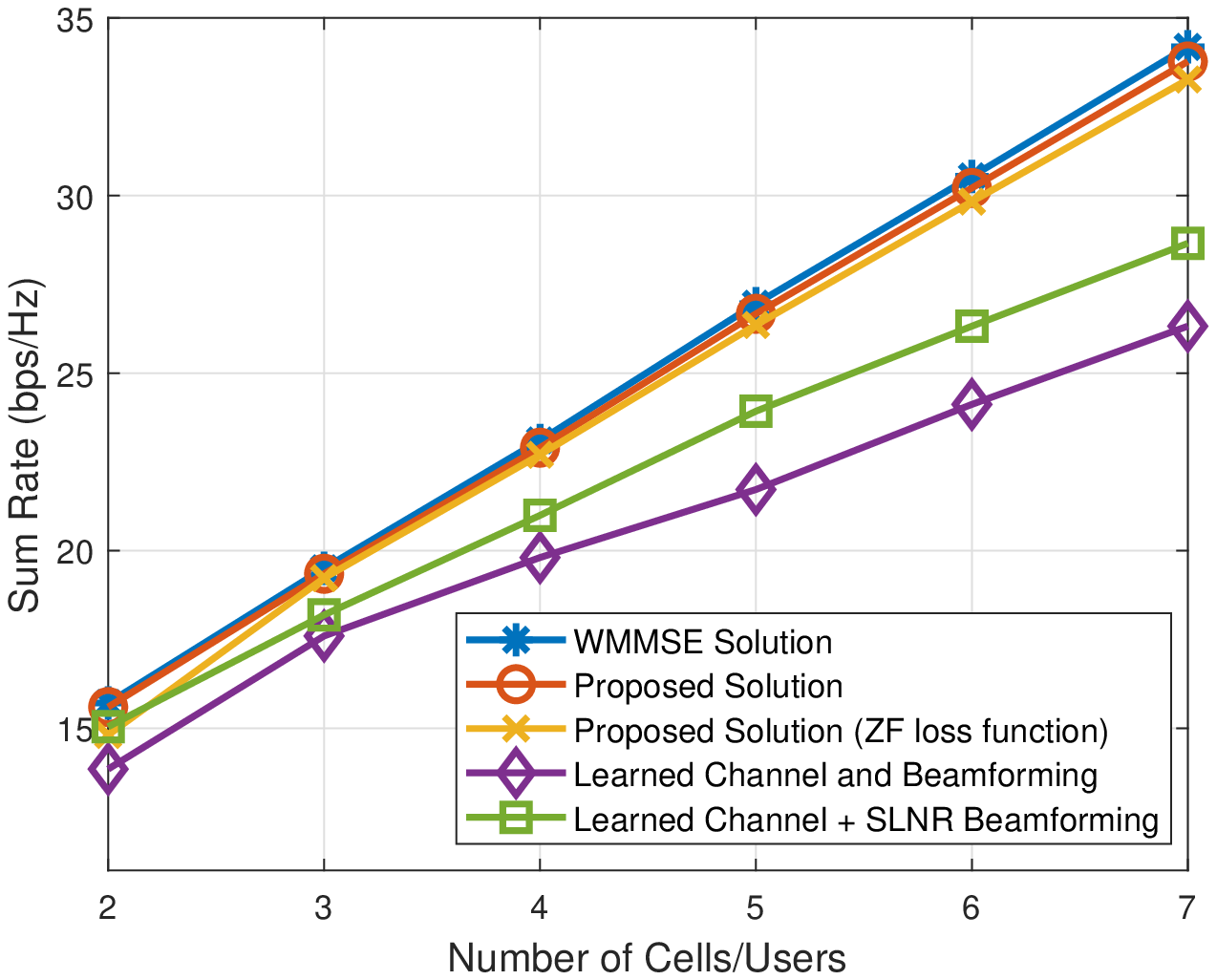}\label{fig User multicell}
	\end{minipage}
}
\centering
\caption{The sum rate performance of the trained model under the generalization evaluation with (a) different total transmit power and (b) numbers of cells.}
\end{figure}

\begin{figure}[t]
\centering
\includegraphics[width=2.5in]{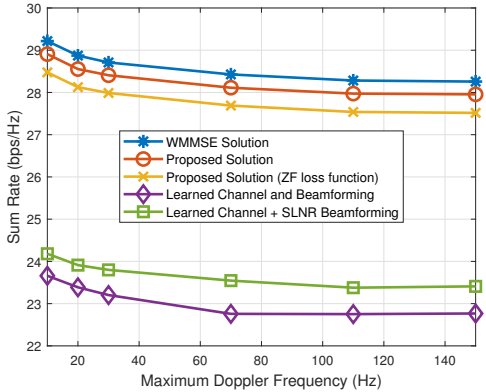}
\caption{The sum rate performance of the trained model under the generalization evaluation with different maximum Doppler frequencies.}
\label{fig doppler multicell}
\end{figure}

In this subsection, we study the generalization performance of the proposed joint training method in the multi-cell massive MIMO scenario when $N_t=32$. The model to be tested is trained  with a total transmit power $P= 10$ dBm  for the originally considered 7-cell scenario and no Doppler effect is considered.
First, we evaluate the performance of the trained model  under different levels of transmit power in Fig. \ref{fig PdBm multicell}. It is seen that  the inference of this model is still capable of achieving a tight performance close to the  WMMSE solution as the transmit power is near 10 dBm, but the performance gap increases as the testing power deviation is large. This shows the proposed method can generalize to   scenarios where there is a small variation of the testing power.
Secondly, we consider the scenario where the system has different number of cells and the results are presented in Fig. \ref{fig User multicell}. It is seen that both   proposed solutions achieve  the sum rate performance close to   the WMMSE solution under different cell numbers, which confirms that the proposed distributed learning solution generalizes well as the number of cells increases. Finally, we consider the user mobility characterized by the Doppler effect in Fig. \ref{fig doppler multicell}. Note that in practical systems, the Doppler effect may be estimated and compensated at the receiver side \cite{zhou_doppler}. Here we assumed the Doppler effect is not compensated so that it will influence the estimated CSI and the sum rate performance. We consider the maximum Doppler frequency to be in the range of 10 to 150 Hz, which corresponds to a speed range from 1  to 20 m/s. Fig. \ref{fig doppler multicell} shows the inference results of the trained model  evaluated by considering different maximum Doppler frequencies. It is seen that the sum rate performance of the proposed solutions shows moderate degradation compared to the WMMSE solution as the Doppler frequency increases,  but  is still significantly higher than the benchmark solutions.

\section{Conclusions}\label{conc}
In this paper, we have proposed a new downlink beamforming optimization algorithm to maximize the sum rate using deep learning when only the uplink channel information is available, but its mapping to the downlink channel is unknown. We introduced a model-driven learning approach by exploiting the structure of the optimal beamforming solution to facilitate an effective neural network design. The proposed approach was extended to   massive MIMO and distributed multicell scenarios. Simulation results demonstrated that our proposed algorithm can approach the performance of the conventional WMMSE algorithm, and achieves a much higher sum rate than the benchmark schemes. These results show the importance of model-driven and holistic learning  approaches to optimize downlink beamforming in practical systems.

\section*{Appendix A}
From \eqref{mse}, the MSE of channel estimation can be expressed as
\bea\label{mse2}
\mbox{MSE}_H &= &\| \qY\qR + \qB - \qH_U\|^2\notag\\
&=&\|  ( \qH_U\qX + \qN)\qR + \qB - \qH_U\|^2\notag\\
&=&\|  \qH_U\qX \qR+  \qN\qR  + \qB - \qH_U\|^2\notag\\
&=&\| \qH_U(\qX \qR-\qI) +  \qN\qR  + \qB \|^2.
\eea

Because $\qH_U = \bar \qH_U + \Delta \qH_U$, where $\Delta \qH_U$ has a zero mean and $\qQ=E[\Delta \qH^H \Delta \qH]$. Then \eqref{mse2} can be further written as
\bea
\mbox{MSE}&=&\|\Delta\qH(\qX \qR-\qI) +  \qN\qR  + \qB + \bar\qH(\qX\qR-\qI)\|^2.
\eea

Apparently the optimal $\qB$ should satisfy
\be\label{B_opt}
\qB^* = -  \bar\qH_U(\qX\qR-\qI).
\ee
Then
\bea\label{mse3}
\mbox{MSE}&=&\mathrm{E}\|\Delta\qH_U(\qX\qR-\qI) +  \qN\qR\|^2\notag\\
&=& \tr[\Delta\qH_U(\qX\qR-\qI)(\qX\qR-\qI)^H)\Delta\qH_U^H] + \tr(\qR^H \qN^H \qN \qR)\notag\\
&=&  \tr[\qR^H(\qX\qQ\qX^H+ N_t \sigma_n^2\qI)\qX - \qX \qR\qQ - \qQ\qR^H\qX^H+\qQ].
\eea

By setting the first order derivative of $\eqref{mse3}$ to be zero, the optimal $\qR$ is given by
\be\label{R_opt}
    \qR^* = \left(\qX\qQ\qX^H+ N_t \sigma_n^2\qI\right)^{-1}\qX^H\qQ.
\ee

The optimal $\qB^*$ and the linear MMSE estimation $\hat\qH_U$ can be obtained by substituting \eqref{R_opt} into \eqref{B_opt} and \eqref{HU}, respectively.

\end{document}